\documentclass[aps,groupedaddress,showpacs,amsfonts,amsmath,floatfix,eqsecnum,nofootinbib
]{revtex4}

\usepackage[usenames]{color}

\usepackage{epsfig}
\usepackage{graphicx}
\usepackage{mathptmx}      
\usepackage{latexsym}
\usepackage{bm}
\usepackage{amstext}\usepackage{array}\usepackage{multirow}

\def\slashchar#1{\setbox0=\hbox{$#1$}
   \dimen0=\wd0 \setbox1=\hbox{/} \dimen1=\wd1
   \ifdim\dimen0>\dimen1 \rlap{\hbox to \dimen0{\hfil/\hfil}} #1
   \else  \rlap{\hbox to \dimen1{\hfil$#1$\hfil}} / \fi}

\def\Tr{{\rm Tr}}

\newcommand{\Eq}[1]{Eq.~(\ref{eq:#1})}

\newcommand{\ignore}[1]{}

\newcommand{\R}{\mathbb{R}}
\newcommand{\C}{\mathbb{C}}
\newcommand{\Z}{\mathbb{Z}}
\renewcommand{\Re}{{\mathrm{Re}}\,}
\renewcommand{\Im}{{\mathrm{Im}}\,}

\newcommand{\fm}{{\phantom{-}}}
\newcommand{\fz}{{\phantom{0}}}

\newcommand{\esp}[1]{{\langle #1 \rangle}}

\begin{document}

\title{Gibbs sampling of complex-valued distributions}

\author{L.L. Salcedo}
\email{salcedo@ugr.es}

\affiliation{Departamento de F\'{\i}sica At\'omica, Molecular y Nuclear and \\
  Instituto Carlos I de F\'{\i}sica Te\'orica y Computacional, \\ Universidad
  de Granada, E-18071 Granada, Spain.}

\date{\today}

\begin{abstract}
A new technique is explored for the Monte Carlo sampling of complex-valued
distributions. The method is based on a heat bath approach where the
conditional probability is replaced by a positive representation of it on the
complex plane. Efficient ways to construct such representations are also
introduced. The performance of the algorithm is tested on small and large
lattices with a $\lambda\phi^4$ theory with quadratic nearest-neighbor complex
coupling. The method works for moderate complex couplings, reproducing
reweighting and complex Langevin results and fulfilling various
Schwinger-Dyson relations.
\end{abstract}

\pacs{ 05.10.Ln, 02.70.-c, 02.70.Ss, 12.38.Gc }
\keywords{}

\maketitle

\tableofcontents

\section{Introduction}

In many physical problems, including statistical mechanics and field theory on
the lattice, one has to deal with a large number of variables.  Simple
estimates show that, when confronted with a generic integral of high
dimensionality, the Monte Carlo method is often the most efficient strategy
\cite{Madras::2002}. In a typical application the configurations $x$ of the
physical system have a probability distribution $P(x)$ in the form of a
Boltzmann weight, $P(x)= e^{-S(x)}$, where $S(x)$ is the action (or the
Hamiltonian) of the configuration, and expectation values of the observables
$A(x)$ come as averages weighted with $P(x)$. The standard approach is then to
produce a sample of the distribution $P(x)$, and obtain an estimate of the
expectation value of the observables from the arithmetic mean of that
sample. Sampling a distribution means to produce points $x$ with a frequency
that in average equals $P(x)$.  In this sense, when the weights of the
distribution are negative or complex a direct sampling becomes meaningless.
Such complex weights do appear in many instances, e.g. in the presence of
fermions \cite{Troyer:2004ge}, in quantum chromodynamics with a baryon number
chemical potential \cite{Hasenfratz:1983ba} or in real time path integral
formulations of quantum mechanics (as opposed to Euclidean time ones). The
impossibility of a direct sampling of the distribution of interest in those
cases constitutes the well known sign (or phase) problem. To face this
situation a number of ideas, specific or generic, have been proposed in the
literature (see for instance
Refs.~\cite{Parisi:1984cs,Hamber:1985qh,Bhanot:1987nv,Salcedo:1993tj%
  ,Kieu:1993gw,Salcedo:1996sa,Barbour:1997ej,Chandrasekharan:1999cm%
  ,Fodor:2001au,Prokofiev:2001,Baeurle:2002,Moreira:2003,Azcoiti:2004ri%
  ,Berges:2005yt,Aarts:2008wh,Banerjee:2010kc,Bloch:2011jx%
  ,Cristoforetti:2013wha,Rota:2015}), yet, at present there is no efficient
approach to deal with the problem of sampling generic complex distributions in
a systematic way. Certainly nothing as universally valid and efficient as,
e.g., the Metropolis algorithm in known for the complex case.

Since a straight sampling of a complex distribution is not possible some
oblique approach is needed. Let us mention just two techniques. One is to
sample a positive auxiliary probability distribution $P_0(x)$, introducing
a compensating weight $P(x)/P_0(x)$ in the observables. This is the
reweighting technique. This method is very general, but it suffers from the
well known overlap problem for large systems
\cite{Barbour:1997ej,Crompton:2001ws}.

The other possibility we mention is based on relaxing the concept of sampling
from configurations to observables. Instead of producing sequences of suitable
distributed {\em points} (configurations) and then compute the observables for
them, it is sufficient to have an algorithm producing, for each observable, an
stochastic sequence of values in such a way that the arithmetic mean of those
sequences reproduce (in average) the correct expectation values. The key point
here is that the random values assigned to the observables need not, and in
general will not, correspond to actual configurations of the system. Since
there is a lot of freedom to do this (for each observable, many different
distributions can be devised having the same expectation value) a practical
way to proceed is by generating {\em complex} configurations (regarded as
basic observables), for which the other observables are computed. This is the
representation technique which is the main focus of this work.

The representation technique relies on constructing a suitable {\em real and
  positive} distribution, $\rho(z)$, defined on the complex plane $\C^d$, {\em
  representing} the original complex probability $P(x)$ defined on
$\R^d$. This means that for any observable $A$, the expectation value of
$A(x)$ with $P(x)$ equals the expectation value of $A(z)$ with $\rho(z)$,
where $A(z)$ stands for the analytic continuation of $A(x)$ from the real to
the complex manifold. Standard importance sampling is then applied to
$\rho(z)$.  

Note that in the representation approach $\esp{1}=1$ automatically. This is in
contrast to reweighting. There, barring the rare cases in which the
normalizations of $P$ and $P_0$ are known, the normalization ensuring
$\esp{1}=1$ has to be enforced a posteriori, using the very Monte Carlo
calculation.

The idea of trading the complex distribution $P(x)$ by a positive
representation of it is implicit in the complex Langevin approach
\cite{Parisi:1980ys,Parisi:1984cs}. Unfortunately the complex Langevin
approach is not the definitive answer to the complex sampling problem. For one
reason, it is limited to actions $S(x)$ having an analytical extension
on the complex plane, and for another, the algorithm is not always convergent,
or even worse, sometimes it converges to wrong equilibrium solutions, that is,
to $\rho(z)$ which are representations of complex distributions different from
the target distribution $P(x)$,
\cite{Ambjorn:1985iw,Ambjorn:1985cv,Salcedo:1993tj,Aarts:2011ax%
  ,Pehlevan:2007eq,Bloch:2015coa}.

As alternatives to the complex Langevin approach, S\"oderberg first considered
representations not directly based on a complex Langevin algorithm
\cite{Soderberg:1987pd}. An explicit formulation and discussion of the concept
of representations by themselves was presented in \cite{Salcedo:1996sa}.
There, explicit representations were constructed for many complex
probabilities, in particular, Gaussian times polynomial of any degree and any
number of dimensions and arbitrary complex distributions with support at a
single point (these are sum of derivatives of Dirac deltas at the same point).
Necessary and sufficient conditions for the existence of positive
representations in $\R^n$ were given in \cite{Weingarten:2002xs}, as well as
the proof of existence for $n=1$. In \cite{Salcedo:2007ji} it was shown that
essentially any complex probability on $\R^n$ admits a positive
representation, as do complex probabilities defined on arbitrary compact Lie
groups (this covers the case of periodic distributions) and explicit
constructive methods were presented. Recent work along the same lines of
representations not based on complex Langevin can be found in
\cite{Wosiek:2015iwl,Wosiek:2015bqg}.

Despite the results in \cite{Salcedo:1996sa,Salcedo:2007ji} the practical
problem of representing and sampling complex distributions is not solved for
two reasons related to locality and uniqueness.

As is well known the Monte Carlo approach is more efficient than other
approaches when the number of variables (degrees of freedom) involved, i.e.,
the number of dimensions of the configuration manifold, is large. However a
good performance of the method requires the action $S(x)$ (and by extension,
the probability) to be {\em local}. By local it is meant that the action is
the sum of terms each of them depending on a few variables and each variable
appearing only in a few such terms.  The requirement of locality is often
essential to have a numerically efficient update procedure of the
configurations in Monte Carlo calculations.  E.g., the coupling between too
many variables is the reason why perfect actions \cite{Bietenholz:1995cy} are
not numerically favored in practice.

Therefore a key question to carry out a Monte Carlo sampling of a complex
probability by means of a representation is whether $\rho(z)$ is local or
not. The great virtue of the complex Langevin approach is that the algorithm
retains the locality of the original complex action. On the contrary, the
constructions of representations found in \cite{Salcedo:1996sa,Salcedo:2007ji}
are non local even if the complex action is. 

The second problem is that of uniqueness. The representation of a given $P(x)$
is not unique. Many different $\rho(z)$ produce the same expectation values on
the set of holomorphic observables \cite{Salcedo:1996sa}. Most such
representations are useless since they go deeply into the complex plane, where
analytically continued observables behave wildly and variances become large.
This implies that in the representation technique (including complex Langevin)
there is an analog of the overlap issue existing in the reweighting approach;
a representation may be formally correct yet produce unacceptable fluctuations
in the Monte Carlo estimates.

In view of these impediments to the direct construction of a representation
for a given complex distribution $P(x)$ in the many-dimensional case, here we
explore a heat bath approach. The update is carried out sequentially for each
of the variables, keeping the remaining variables fixed to their current
value. To do the update of a variable, we replace its conditional probability,
a complex function, by a positive representation of it on the complex
plane. The point of following this procedure is that, being the conditional
probability a function of a single variable, obtaining a representation for it
is relatively easy and inexpensive. The locality issue is bypassed since, if
the complex action is local, the representation of the conditional probability
will also depend locally on the remaining variables. The quality of the
representation, regarding variances, can also be controlled more easily in the
one-dimensional case.

Nevertheless, in the complex case we are not protected by the standard
convergence theorems for Markov chains based on positive probabilities
\cite{Madras::2002}, and this can lead to difficulties, as already found in
the complex Langevin approach.  That problems may arise in a complex heat bath
approach can be understood from the following consideration. Assume
$P(x_1,x_2)$ is the complex probability to be sampled.  One would be safe by
constructing the positive two complex-dimensional representation
$\rho(z_1,z_2)$. However, if a Gibbs approach is used, one needs to represent
instead the conditional probability $P(x_1|z_2) = P(x_1,z_2)/P(z_2)$, where
the marginal probability $P(z_2)= \int dx_1 P(x_1,z_2)$ is required at complex
values of $z_2$. The trouble is that, even if $P(x_2)$ is never zero on the
real axis, it can have zeros on the complex plane.

In any case, in our view, presently lacking reliable and general sampling
methods of complex probabilities, it seems worthwhile to explore and test new
approaches to assess their performance. This work is organized as follows. In
Sec. \ref{sec:2} we discuss the concept of representation (Sec. \ref{sec:2.A})
and show that the complexness of $P(x)$ puts restrictions on the how localized
the representation can be on the complex plane, i.e. on the quality of the
possible representations (Sec. \ref{sec:2.B}). Also, in Sec. \ref{sec:2.C}
constructive representation techniques are presented, where the quality of the
representations can be optimized, both in one- and in higher
dimensions. Sec. \ref{sec:3} is devoted to discuss and analyze the complex
heat bath approach. The method is introduced in Sec. \ref{sec:3.A} and it is
applied to a quadratic action on a hypercubic lattice in Sec. \ref{sec:3.B}. A
deeper study of the performance of the algorithm is presented in
Sec. \ref{sec:3.C}. There a $\lambda \phi^4$ theory with complex
nearest-neighbor complex coupling is analyzed. The Monte Carlo results of the
complex heat bath algorithm are compared to those obtained from reweighting
and complex Langevin. Conclusions are presented in Sec. \ref{sec:4}

\section{Complex probabilities and representations}
\label{sec:2}

\subsection{Concept of representation of a complex probability}
\label{sec:2.A}

We will call {\em complex probability} to any normalizable complex
distribution $P(x)$. After normalization, it make take negative or complex
values instead of being real and positive for all $x$. The expectation values
of observables $A(x)$ are obtained as usual through
\begin{equation}
\esp{A}_P = \frac{\int d\mu(x) \, P(x) A(x)}{\int d\mu(x) \,P(x)}
.
\end{equation}
Here $d\mu(x)$ is the appropriate positive measure in the $x$-manifold. In
what follows we often assume $x\in\R^n$, or a periodic version of
$\R^n$, and $d\mu(x)=d^nx$.

A {\em representation} of $P$ is an ordinary probability $\rho(z)$ (i.e., a
normalizable, real and positive distribution) defined on the complexified
manifold, $\C^n$, and such that it produces the same expectation values as $P$
upon analytical extension of the observable, that is,
\begin{equation}
\esp{ A(x) }_P = \esp{ A(z) }_\rho
,
\label{eq:2.2}
\end{equation}
where
$A(z)$ refers to the analytical extension of $A(x)$ from $\R^n$ to $\C^n$, and
\begin{equation}
\esp{A}_\rho = \frac{\int d^{2n}z \, \rho(z) A(z)}{\int d^{2n}z \,\rho(z)}
.
\end{equation}

We will often refer to the real manifold as the {\em real axis} of $\C^n$.  In
the periodic case $\rho(z)$ is normalized as (assuming a $2\pi$ period in each
direction)
\begin{equation}
\int_{[0,2\pi]^n} d^n x \,
\int_{\R^n}d^n y \, \rho(x+iy) = 1
,
\end{equation}
so the complexified manifold is non compact in the imaginary direction.

Strictly speaking, \Eq{2.2} expresses that $\rho$ is a representation of $P$
valid for the observable $A$. Generically we will say that $\rho$ is a
representation of $P$ when it is valid for a large set of sufficiently well
behaved observables or test functions. One can specify the set of test
functions $A$ as that of polynomials of $z$, or that of plane waves,
$\exp(-ikz)$, $k\in\R^n$. Since the plane waves grow exponentially in the
complex plane the latter set is more restrictive in general, nevertheless, we show
below that often the support of $\rho$ can be chosen to remain within a finite
strip along the real axis. This avoids any problem related to the exponential
growth of $A(z)$. In the case of periodic distributions the set of Fourier
modes $\exp(-ikz)$, $k\in\Z^n$ is a natural choice.

If the support of a representation fills $\C^n$, observables with singular
points cannot be reproduced by it. However, representations are constructed
below with support smaller than $\C^n$.  In that case some observables with
singularities on the complex plane can also be reproduced. This is guaranteed
when the singularities lie outside some simply connected region containing
both the support of the representation and the real axis.  A one-dimensional
example is $P(x)=\exp(-(x-i)^2/2)/\sqrt{2\pi}$, which can be represented by
$\rho(x,y)=\delta(y-1)\exp(-x^2/2)/\sqrt{2\pi}$. This representation correctly
reproduces $A(x)=1/(x-z_0)$ provided $\Im z_0$ is either negative or larger than
one.

\subsection{Conditions on the support of the representations}
\label{sec:2.B}

The application of Monte Carlo sampling to a complex distribution, by means of
a representation of it, differs in an essential aspect from the standard case
of positive distributions. In the standard case, one often makes a direct
sampling of the distribution. Rarely a reweighting method is used since this
tends to increase the variance in the expectation values (for generic
observables).
In the complex case one has to somehow construct the
representation and such representation is by no means unique.  Any target
complex probability $P(x)$ admits many valid representations, of all them
formally correct but vastly different as regards to performance.  The non
uniqueness implies in particular that the expectation value of non holomorphic
observables, $B(z,z^*)$, can be different for different representations. An
estimate of the variance, e.g., $B=|A|^2-|\esp{A}|^2$ (where $A(z)$ is
holomorphic), follows this rule and so, while $\esp{A}$ does not depend on the
representation, its Monte Carlo estimate does.

For instance, if $\rho(z)$ is a representation, it is easy to show that the
new distribution $\rho^\prime(z)$ obtained by convolution of $\rho$ with any
positive distribution of the type $C(|z|^2)$ (i.e., rotationally symmetric)
provides a new representation \cite{Salcedo:1996sa}.\footnote{Many more new
  representations can be obtained by adding a {\em null representation},
  $\rho_0$, i.e. one with vanishing expectation value for any holomorphic
  observable, however, there is no guarantee that the sum will be positive
  definite.} This new representation will be wider, more spatially extended,
than the previous one, and this is an undesirable feature. In general one will
want a representation $\rho(z)$ as localized and close to the real axis as
possible. The reason is that most observables will grow, often exponentially,
as one departs from the real axis, and as a consequence the statistical
fluctuations also grow, rendering a Monte Carlo approach less efficient.

The overlap problem of the reweighting technique for positive distributions
exists, to some extent, also in the representation technique. In principle,
the best representation is the most localized one, in the sense that variances
of observables will be smaller. Incidentally, one sensible way to measure the
localization is through the (relative) entropy,\footnote{In \Eq{ent},
  $\rho_{{\rm ref}}$ is some fixed reference distribution. $\rho$ and
  $\rho_{{\rm ref}}$ are positive and normalized.}
\begin{equation}
{\mathcal S}(\rho) = \esp{-\log(\rho/\rho_{{\rm ref}})}_\rho
,
\label{eq:ent}
\end{equation}
rather than the variance matrix of $z$. To see this consider a $\rho(z)$
composed of two well-separated narrow Gaussian functions.  The quantity
$\esp{|z-\esp{z}|^2}_\rho$ measures the typical separation between points in
the support of $\rho(z)$ and this will be large if the two Gaussian functions
are far from each other. On the other hand the entropy is independent of the
separation (as long as the two Gaussians have a negligible overlap). One can
obtain the expectation values of observables for each of the Gaussians with
small variance (since they are narrow) and then combine the result with a
final small error. As said, proximity to the real axis is convenient too,
however, for a valid representation this cannot be controlled since it is
dictated by $P(x)$ as we argue below.

From the point of view of the localization, the best choice would be $\rho(z)=
P(x) \delta(y)$, where $y$ refers to the imaginary axis
coordinate. Unfortunately such representation is not positive for complex
$P$. In fact, there is a kind of uncertainty principle implying that the less
positive $P$ is, the wider $\rho$ should be on the complex plane. The width
here refers to the extension of the support of $\rho$ in the imaginary
direction.

To make this principle more precise let us consider the one-dimensional case,
$x\in\R$, although similar considerations hold for any number of dimensions.
Let $P$ and $\rho$ be normalized, and
\begin{equation}
{\tilde P}(k) := \int dx e^{-ikx} P(x) = \esp{e^{-i k x}}_P
\end{equation}
then, using $z=x+iy$,
\begin{equation}
|{\tilde P}(k)| = |\esp{e^{-ikx}}_P| = |\esp{e^{-i k z}}_\rho|
\le  |\esp{|e^{-i k z}|}_\rho|
= \esp{e^{ k y}}_\rho
.
\label{eq:6}
\end{equation}
Therefore, if the support of $\rho$ lies in the region $y\le Y_1$, it follows
that $\esp{e^{ k y}}_\rho \le e^{k Y_1}$ for all positive $k$, and hence
$|{\tilde P}(k)| \le e^{k Y_1}$ ~$\forall k>0$.  Likewise, if the support of
$\rho$ lies in $y \ge Y_2$, necessarily $|{\tilde P}(k)| \le e^{k Y_2}$ for
all negative $k$. It follows that
\begin{equation}
Y_1 \ge Y_+ \equiv \max_{k>0}\left(\frac{1}{k}\log|\tilde{P}(k)|\right)
,
\quad
Y_2 \le Y_- \equiv \min_{k<0}\left(\frac{1}{k}\log|\tilde{P}(k)|\right)
.
\label{eq:2.7a}
\end{equation}
In other words, any representation $\rho$ must have some support in the region
$y \ge Y_+$ and as well as in the region $y \le Y_-$. If the support of $\rho$
falls in a strip $Y_2 \le y \le Y_1$ the width of the strip is restricted by
the conditions $Y_1 \ge Y_+$ and $Y_2 \le Y_-$.  These formulas extend
immediately to the case of periodic distributions.

As an example, consider the one dimensional action
\begin{equation}
 S(x)= \frac{\beta}{4} x^4 + i q x
,\qquad
\beta > 0.
\label{eq:2.7}
\end{equation}
For $\beta=0.5$ and $q=2$ one finds $Y_+=4.29$. Thus a proper representation
of this complex action must have some support above $y=4.29$.  As it turns
out, the complex Langevin algorithm applied to this action with $q$ positive
will produce an equilibrium distribution entirely located {\em below} the real
axis. This is because once the random walk goes below the real axis it can
never get above it.\footnote{We always refer to the standard implementation of
  the complex Langevin method which has only horizontal noise and with no
  kernel.} Therefore we know, without doing the actual calculation, that the
complex Langevin algorithm converges to the wrong equilibrium distribution in
this case. Actually, this action is not at all pathological and it admits a
valid representation of the two-branches type described below, in Section
\ref{sec:II.C.3}.

These bounds can be easily generalized to other observables and any number of
dimensions as follows: Let $P(x)$ be a complex probability in $\R^n$ and
$\rho(z)$ a representation of $P$, and let $A(z)$ be a test function. Then
\begin{equation}
|\esp{A(x)}_P| = |\esp{A(z)}_\rho| \le
\max\{ |A(z)|, ~ z\in{\rm supp}(\rho)  \}
.
\end{equation}
This inequality puts conditions on the support of $\rho$. Indeed, given an
observable $A$, its expectation value $a$ does not depend on the choice of
$\rho$. Then one can define the set $\mathcal{A}$ of points where $|A(z)| \ge
|a|$.  The inequality implies that the support of any proper representation of
$P$ must have some overlap with $\mathcal A$, and this for all test functions
$A$.

For instance, again for the action of \Eq{2.7} with $\beta=0.5$ and $q=2$, one
finds that $\esp{1/(x-i)} = -2.82 i$. Thus, the set $\mathcal{A}$ is the disk
of radius $0.35$ centered at $z=i$. Since the lower half-plane has no overlap
with this disk, any representation without some support above the real axis
can automatically be ruled out.\footnote{A tricky point is that the argument
  works because the singularity at $z=i$ would lie beyond the support of any
  such representation and so for them $1/(x-i)$ would be an acceptable test
  function.  The two-branches representations (discussed below) for this
  action have support above the singularity and so for them $1/(x-i)$ is not
  an acceptable test function, and in fact, their support has no overlap with
  the disk $|z-i|<0.35$.}

\subsection{Construction of explicit representations}
\label{sec:2.C}

Barring those of \cite{Salcedo:1996sa,Salcedo:2007ji}, the representations of
complex probabilities existing in the literature are limited to a few cases:
quadratic actions, which can be obtained analytically, complex Langevin
constructions, which have only an empirical basis, and some representations
constructed by first choosing a suitable $\rho(z)$ and then finding to which
complex probability it corresponds. This can be done, e.g., by means of the
projection \cite{Salcedo:1993tj}
\begin{equation}
\esp{A(z)}_\rho = \int d^nxd^ny\, \rho (x,y) A(x+iy)
=
\int d^nx \, \left(\int d^n y \, \rho(x-iy,y)\right) A(x)
=:
\int d^nx \, P(x) A(x) =  \esp{A(x)}_P
.
\end{equation}

In this section we discuss new some explicit constructions of representations
for generic complex probabilities. Further constructions, including the case
of complex probabilities defined on a compact Lie group can be found in
\cite{Salcedo:2007ji}.

Note that the need of an analytical extension of $P(x)$ itself (in addition to
that of the observables) is not a general requirement to have a
representation.  It is an idiosyncracy of the complex Langevin algorithm
(which actually requires a holomorphic $\log P(x)$) and of some other
approaches (e.g. the one-branch representations, and the complex Gibbs
sampling discussed below) while other methods apply to generic normalizable
complex distributions.

\subsubsection{An explicit representation in one dimension}

The one-dimensional distribution 
\begin{equation}
Q(x)=\delta(x)+\delta^\prime(x)
\end{equation}
admits the representation
\begin{equation}
q(z)= \frac{1}{8\pi}\left|1-\frac{z}{2}\right|^2 e^{-|z|^2/4}
\label{eq:2.1}
\end{equation}
as is readily verified by checking that $\esp{ x^n}_Q = \esp{z^n}_q$ for all
non negative integer $n$. 

This basic representation can then be used to construct representations of
arbitrary complex probabilities in any number of dimensions
\cite{Salcedo:2007ji}. Here we show this for a one-dimensional $P(x)$ which is
assumed to be normalized. Let us decompose $P(x)$ as
\begin{equation}
P(x) = P_0(x) + F^\prime(x)
,
\qquad
P_0(x) > 0, \quad \int dx \, P_0(x) = 1
. 
\end{equation}
$P_0(x)$ can be chosen in many ways and this choice fixes $F(x)$: Since both
$P$ and $P_0$ are normalized, the function 
\begin{equation}
F(x)= \int_{-\infty}^x \left( P(x^\prime) - P_0(x^\prime) \right) dx^\prime
\end{equation}
vanishes for large $x$. This $F$ can be written as $F(x)=P_0(x)h(x)$ where
$h(x)$ will be a complex function in general. In this case, a representation
of $P(x)$ is provided by
\begin{equation}
\rho(z) 
= \int dx  \,d^2 z^\prime  \, P_0(x) q(z^\prime) \delta(z-x - h(x)z^\prime)
= \int dx \, P_0(x) \frac{q\left((z-x)/h(x)\right)}{|h(x)|^2}
.
\label{eq:10}
\end{equation}
($\delta(z-z_0)$ refers to the two-dimensional delta on the complex plane.)
To verify that this $\rho(z)$ is really a normalized representation of $P(x)$
let us apply it to a generic observable (using the first form in \Eq{10})
\begin{equation}
\esp{A}_\rho = 
\int d x  \, d^2 z^\prime \, P_0(x) q(z^\prime) \, A(x + h(x) z^\prime)
.
\label{eq:11}
\end{equation}
Now, because $q(z)$ is a representation of $Q(x)$ and $A(z)$ is analytical, it
follows that
\begin{equation}
\esp{A}_\rho = 
\int d x \, dx^\prime \, P_0(x) Q(x^\prime)  \, A(x + h(x) x^\prime)
=
\int d x \, P_0(x) (A(x) - h(x) A^\prime(x))
= \int d x \, P(x) A(x)
= \esp{A(x)}_P
.
\end{equation}
(where $ A^\prime(x)$ denotes the derivative of $A(x)$).

The formula in \Eq{11} already indicates how to carry out a sampling of the
complex probability $P(x)$, namely, sample $x$ with $P_0(x)$ and $z$ with
$q(z)$ and average the values of the observable computed at $x + h(x) z$.

\subsubsection{One-branch representations in one dimension}

The Monte Carlo method suggested in \Eq{11} can be extended to any number of
dimensions, however, the determination of the function $\vec{h}(x)$ (which is
far from unique) is not so straightforward as in the one-dimensional case
\cite{Salcedo:2007ji}. Even in the one-dimensional case the construction
presented above has the problem that the support of $\rho(z)$ will be
spatially more extended than necessary. This problem is common to constructive
approaches of generic type. Essentially, the procedure to obtain \Eq{10} has
been to rewrite $P(x)$ in the form
\begin{equation}
P(x) = P_0(x) + (P_0(x)h(x))^\prime
= \int dx^\prime \, P_0(x^\prime) \left( 
\delta(x^\prime-x) - h(x^\prime) \delta^\prime(x^\prime-x)
\right)
\label{eq:2.19}
\end{equation}
and proceed to replace $\delta(x^\prime-x) - h(x^\prime)
\delta^\prime(x^\prime-x)$ by a representation of it. The width of the
representation of $\delta(x)-h\delta^\prime(x)$ on the complex plane increases
with $|h|$, with a coefficient which depends on the phase of $h$. In the worst
cases, $Y_\pm=\pm|h|$, for $h=\pm i |h|$, respectively ($Y_\pm$ were
introduced in \Eq{2.7a}).

As a rule, the approaches such as that in \Eq{2.19}, based on writing the
target complex probability $P$ as
\begin{equation}
P(x) = \sum_n w_n p_n(x),
\qquad
\rho(z) = \sum_n w_n \rho_n(z),
\end{equation}
where $p_n(x)$ is some fixed basis of normalized complex probabilities with
known representations $\rho_n(z)$ and the $w_n$ are some positive weights,
allow for an immediate representation of $P(x)$ but the width of $\rho(z)$
along the real axis will be fixed by that of the $\rho_n(z)$ (in fact, by the
worst case), so in general $\rho(z)$ will be wider than necessary.

As it turns out, better representations, with width closer to the bounds
$Y_\pm$ discussed above, can be obtained by adapting the representation to the
target complex probability $P(x)$. An obvious example is a probability of the
type
\begin{equation}
P(x) = P_0(x-a)
\label{eq:2.18}
\end{equation}
where $P_0(x)$ is positive (for real $x$) and the constant $a=a_R+i a_I$ may
be complex. In this case $\rho(z) = P_0(x-a_R)\delta(y-a_I)$ ($z=x+i y$) is a
valid representation, and in fact the best one. The support lies on a straight
line parallel to the real axis. We will refer to representations with support
on a single line as one-branch representations.\footnote{Of course, if $P(x)$
  happens to have the form in \Eq{2.18} the representation $\rho(z) =
  P_0(x-a_R)\delta(y-a_I)$ works in any number of dimensions.}

\begin{figure}[h]
\begin{center}
\epsfig{figure=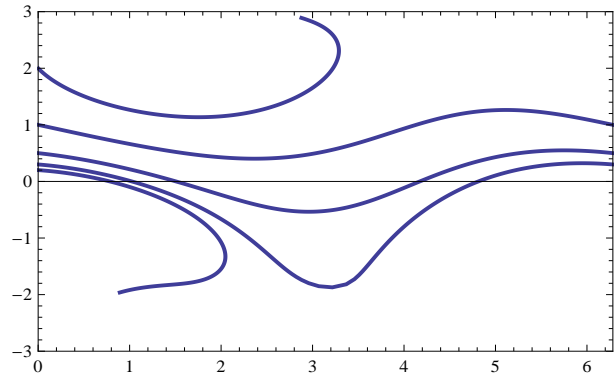,height=50mm,width=80mm}
\end{center}
\caption{One-branch paths on the complex plane $\phi$, parameterized by $0\le
  u \le 1$, for the periodic complex probability with action $S(\phi) = a
  e^{i\phi} + b e^{-i\phi}$, with $a=0.5$ and $b=0.15i$. From bottom to top,
  the paths start at $\phi= 0.2i, 0.3i, 0.5i, i, 2i$. The three middle paths
  are periodic and so any of them provides a one-branch representation of the
  complex probability.}
\label{fig:1branch}
\end{figure}

Not all complex probabilities admit a one-branch representation. For $P(x)$
defined on $\R$, this requires the existence of a path $z(t)$ on the complex
plane connecting $x=\pm\infty$, thus ensuring that integration along the path
is equivalent to integration along the real axis, such that $P(z)dz$ stays
positive, ensuring a positive weight \cite{Soderberg:1987pd}. In this case
$P(x)$ can be sampled by using a standard inversion method, namely, for $u\sim
{\rm U}(0,1)$, the value of $z$ is determined by the condition $u =
\int_{-\infty}^zP(z^\prime)dz^\prime$. For periodic $P(x)$, the path must also
be periodic. An example is shown in Fig. \ref{fig:1branch} for $P(\phi) = N
e^{-a e^{i\phi}-b e^{-i\phi}}$ with $a=0.5$ and $b=0.15i$.

Although one-branch representations do not always exist for a given $P(x)$
a possible device is to split $P(x)$ as $P = P_1 + P_2$ with $P_1$ chosen
in such a way that a representation is known for it and $P_2=P-P_1$ admits a
one-branch representation. The idea here is that a common $P_1$ can be used
for a family of $P$, and only the concrete one-branch representation
of $P_2$ has to be constructed in each case, e.g, by integration of
$dz/du=1/P_2(z)$.

\subsubsection{Two-branches representations in one dimension}
\label{sec:II.C.3}

Representations with support on two lines, called here two-branches
representations, exist quite generally for periodic and non periodic complex
probabilities. The ones we consider here are of the type
\begin{equation}
\rho(z) = Q_1(x)\delta(y-Y_1) + Q_2(x)\delta(y-Y_2)
,
\qquad
Q_{1,2}(x) \ge 0
\qquad
Y_1 \ge Y_2
,
\label{eq:15}
\end{equation}
so the support is along the lines $y=Y_1$ and $y=Y_2$ parallel to the real
axis. Here $Q_1(x)$ and $Q_2(x)$ are two suitable positive functions. For
definiteness we have chosen $Y_1 \ge Y_2$. Note that requiring this $\rho(z)$
to be a representation of $P(x)$ is equivalent to imposing the relation
\begin{equation}
P(x) = Q_1(x-iY_1) + Q_2(x-iY_2)
,
\label{eq:15b}
\end{equation}
upon analytical extension of functions $Q_{1,2}(z)$ which are positive on the
real axis. Indeed,
\begin{equation}
\int d^2 z \, \rho(z) A(z) = \int dx \, \left( Q_1(x) A(x + i Y_1 ) + Q_2(x) A(x
+ i Y_2)\right) = \int dx \, \left( Q_1(x-iY_1) + Q_2(x - i Y_2)\right) A(x)
.
\end{equation}

In fact for two given distinct $Y_{1,2}$, the {\em real} functions $Q_{1,2}$
are unique in the non compact case. For periodic probability distributions the
only ambiguity is an additive constant which can be moved between the two
functions.

Let us first analyze the non compact case, that is, $P(x)$ defined on $\R$. We
further assume $P(x)$ to be normalized. Taking a Fourier transform
\begin{equation}
\tilde{P}(k) = \int dx \, P(x) e^{-ikx} = \int d^2z \, \rho(z) e^{-ikz}
=
\tilde{Q}_1(k) e^{kY_1} + \tilde{Q}_2(k) e^{kY_2}
.
\end{equation}
Imposing now the condition that $Q_{1,2}(x)$ are real, and hence
$\tilde{Q}_{1,2}^*(k) = \tilde{Q}_{1,2}(-k)$, one obtains
\begin{equation}
\tilde{Q}_1(k) = \frac{
e^{- k Y_2} \tilde{P}(k) - e^{k Y_2} \tilde{P}^*(-k) }{2 \sinh(k(Y_1-Y_2))} 
,\qquad
\tilde{Q}_2(k) = \frac{
e^{- k Y_1} \tilde{P}(k) - e^{k Y_1} \tilde{P}^*(-k) }{2 \sinh(k(Y_2-Y_1))} 
.
\label{eq:2.25}
\end{equation}
The functions $\tilde{Q}_{1,2}(k)$ are well behaved at $k=0$ since
$\tilde{P}(0)=1$ (any real normalization would do as well) and
\begin{equation}
\tilde{Q}_1(0) = \frac{ \Im\!\esp{x} - Y_2 }{Y_1-Y_2}
,
\qquad
\tilde{Q}_2(0) = \frac{ \Im\!\esp{x} - Y_1 }{Y_2-Y_1}
.
\end{equation}

So a real solution exists and is unique.  An obvious necessary condition for
$Q_{1,2}(x)$ to be non negative is $\tilde{Q}_{1,2}(0) \ge 0$, hence
\begin{equation}
Y_2 \le \Im\!\esp{x} \le Y_1
.
\label{eq:2.27}
\end{equation}
Once again this relation shows that a complex $P(x)$ requires a representation
with a minimum width around the real axis.\footnote{This relation is just the
  condition $Y_1 \ge \log(|{\tilde P}(k)|)/k$ for small positive $k$, and
similarly for $Y_2$, so it is weaker than \Eq{2.7a}.}

The formulas of $\tilde{Q}_{1,2}(k)$ can be reexpressed as convolutions in
$x$-space,
\begin{equation}
Q_1(x) = - 
\frac{1}{2(Y_1-Y_2)}\Im \left(
\tanh\left(\frac{\pi(x+iY_2)}{2(Y_1-Y_2)}\right) * P(x)
\right)
,\qquad
Q_2(x) =  
\frac{1}{2(Y_1-Y_2)}\Im \left(
\tanh\left(\frac{\pi(x+iY_1)}{2(Y_1-Y_2)}\right) * P(x)
\right)
.
\label{eq:2.29a}
\end{equation}
(These formulas rely on the assumption $Y_1>Y_2$.)  Note that, despite the
presence of $\tanh$, the functions $Q_{1,2}(x)$ vanish for large $x$ thanks to
the condition $\int \! dx \, \Im P(x) = 0$.

The construction guarantees that the functions $Q_{1,2}(x)$ are real.
Although lacking a detailed proof (the proof exists for the compact case,
below), one empirically finds that $Y_{1,2}$ can be chosen so that these
functions become non negative. This happens for $Y_1$ and $-Y_2$ sufficiently
large.  Obviously, from the our discussion in Sec. \ref{sec:2.B}, it follows
that this requires $Y_1 \ge Y_+$ and $Y_2 \le Y_-$ (and hence also the weaker
condition in \Eq{2.27}).

We have not observed a significant improvement by optimizing $Y_{1,2}$
separately, so for simplicity we will adopt the symmetric choice
\begin{equation}
Y \equiv Y_1 = -Y_2  > 0
.
\label{eq:2.29}
\end{equation}
Except in the trivial case of $P$ positive, $Q_1$ and/or $Q_2$ have negative
regions for too small values of $Y$.  What is found is that as $Y$ grows, the
minima of $Q_1$ and $Q_2$ grow as well until $Y$ reaches a critical
value. From then on $Q_{1,2}(x)$ are non negative and the minima jump to
$x=\pm\infty$ where these functions vanish. So the critical value of $Y$ is
such that $Q_{1,2}\ge 0$ and $Q_1$ or $Q_2$ vanish at some finite point.

From the numerical point of view the critical value of $Y$ is the optimal one,
since, as a rule, the closer the support of $\rho$ to the real axis the
smaller the variance in the observables. This rule holds in all approaches
based on representing complex probabilities, including complex Langevin.  In
the two-branches case, probabilities which require relatively large values of
$Y$ can be considered as numerically hard, while those admitting small values
are soft.

\begin{figure}[h]
\begin{center}
\epsfig{figure=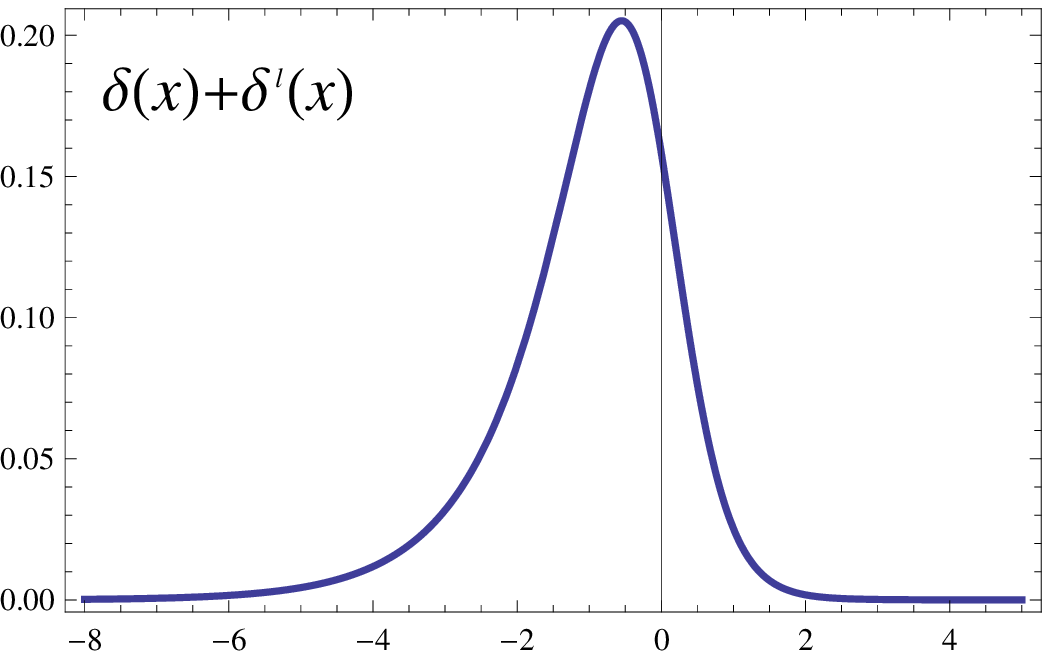,height=50mm,width=80mm}
\epsfig{figure=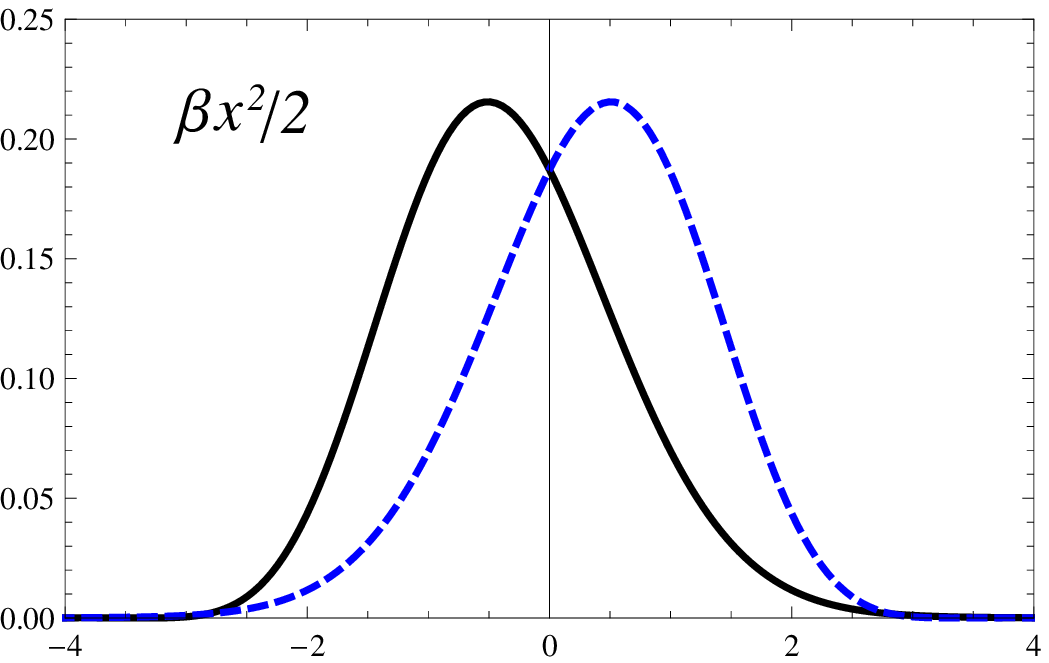,height=50mm,width=80mm}
\end{center}
\caption{Left: Function $Q_1(x)=Q_2(x)$ for
  $P(x)=\delta(x)+\delta^\prime(x)$, using the optimal value $Y=1.58$.  Right:
  Functions $Q_1(x)$ (solid line) and $Q_2(x)=Q_1(-x)$ (dashed line) for the
  two-branches representation of the action $S(x) = \beta x^2/2$ with $\beta=
  1+i$ using the optimal value $Y=0.7$. }
\label{fig:2branchG}
\end{figure}
Examples of two-branches representations are displayed in
Fig. \ref{fig:2branchG}. One such example is the complex action $S(x) = \beta
x^2/2$ with $\beta= 1+i$ using $Y=0.7$. Another is
$P(x)=\delta(x)+\delta^\prime(x)$, using the optimal value $Y=1.58$. Of
course, in this latter case the convolutions in \Eq{2.29a} can be obtained
analytically. The size of the representation $q(z)$ in \Eq{2.1} can be
estimated from $\esp{|z|^2}_q = 6$. This is larger than that of the
two-branches one $\esp{|z|^2}_\rho \approx 5$.  Actually in the two-branches
case $y$ is fixed to known values, either $Y$ or $-Y$, with equal weight, so
this variable does not add to the variance. In this view, recalling our
discussion on the entropy in Sec. \ref{sec:2.B}, the true uncertainty
(understood as uncontrolled fluctuation) would be better estimated from
$\esp{x^2}_{Q_1} \approx 2.5$. This is particularly clear in the present
example due to $Q_1=Q_2$, because $\esp{A} = \esp{A(x+iY)+A(x-iY)}_{Q_1}$
exactly, with no added fluctuation from $y$.

Let us turn now to the case of a one-dimensional {\em periodic} complex
probability $P(x)$. We assume $P$ to be normalized and with period $2\pi$,
$\int_0^{2\pi} P(x) \, dx = 1$. In order to construct a two-branches
representation, Eqs. (\ref{eq:15}) and (\ref{eq:15b}) apply. Decomposing
in Fourier modes
\begin{equation}
P(x) = \frac{1}{2\pi} \sum_{k\in\Z}  e^{ikx} \tilde{P}_k
,
\qquad 
\tilde{P}_0=1,
\end{equation}
and requiring $Q_{1,2}(x)$ to be real periodic functions, one obtains
\begin{equation}
\tilde{Q}_{1,k} = \frac{
e^{- k Y_2} \tilde{P}_k - e^{k Y_2} \tilde{P}^*_{-k} }{2 \sinh(k(Y_1-Y_2))} 
,\qquad
\tilde{Q}_{2,k} = \frac{
e^{- k Y_1} \tilde{P}_k - e^{k Y_1} \tilde{P}^*_{-k} }{2 \sinh(k(Y_2-Y_1))} 
,
\qquad k \not =0
.
\end{equation}

In the periodic case the zero modes of $\tilde{Q}_{1,2}$ are not determined by
$P(x)$, instead one has only the conditions
\begin{equation}
\tilde{Q}_{1,0} + \tilde{Q}_{2,0} = 1,
\qquad
\tilde{Q}_{1,0},\, \tilde{Q}_{2,0} \ge 0
.
\end{equation}
Let $\hat{Q}_{1,2}(x)$ denote the functions $Q_{1,2}(x)$ (reconstructed from
their Fourier modes) without including the zero mode, i.e., $\hat{Q}_r =
Q_r-\tilde{Q}_{r,0}/(2\pi)$, $r=1,2$. It is readily shown that whenever the
following conditions are met
\begin{equation}
\min_x{\hat{Q}_1} \ge -1 , \quad
\min_x{\hat{Q}_2} \ge -1 , \quad
\min_x{\hat{Q}_1}+\min_x{\hat{Q}_2} \ge -1 ,
\label{eq:2.33a}
\end{equation}
(by choosing $Y_{1,2}$
sufficiently large) suitable $\tilde{Q}_{1,0}$, $\tilde{Q}_{2,0}$ can be added
so that $Q_{1,2}(x)$ are non negative, thus providing a two-branches
representation.

\begin{figure}[h]
\begin{center}
\epsfig{figure=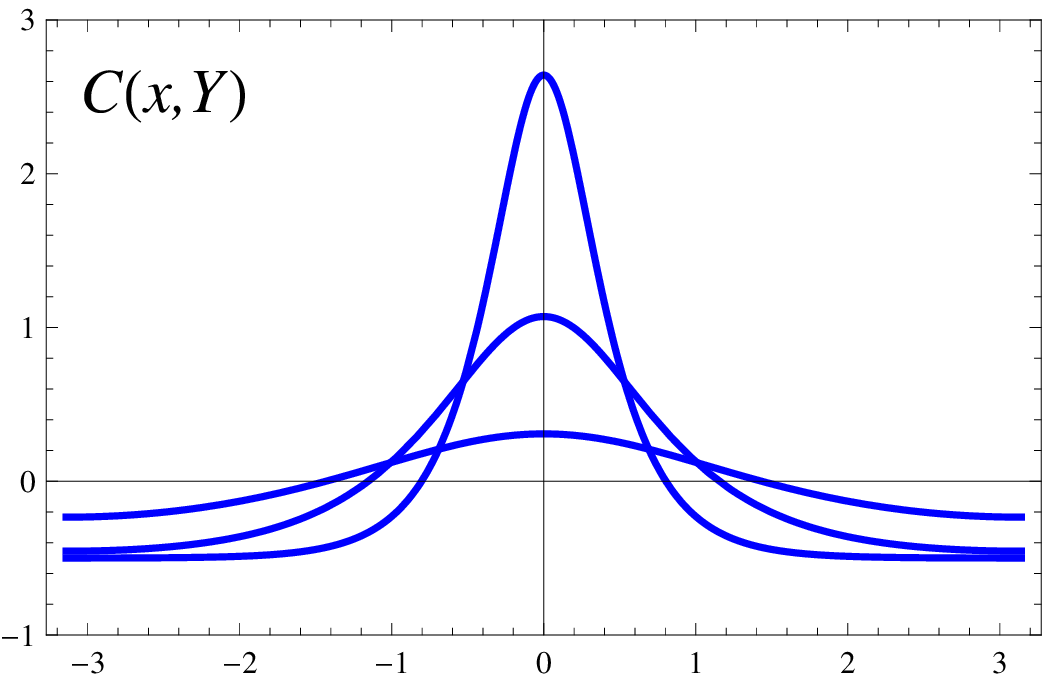,height=50mm,width=80mm}
\epsfig{figure=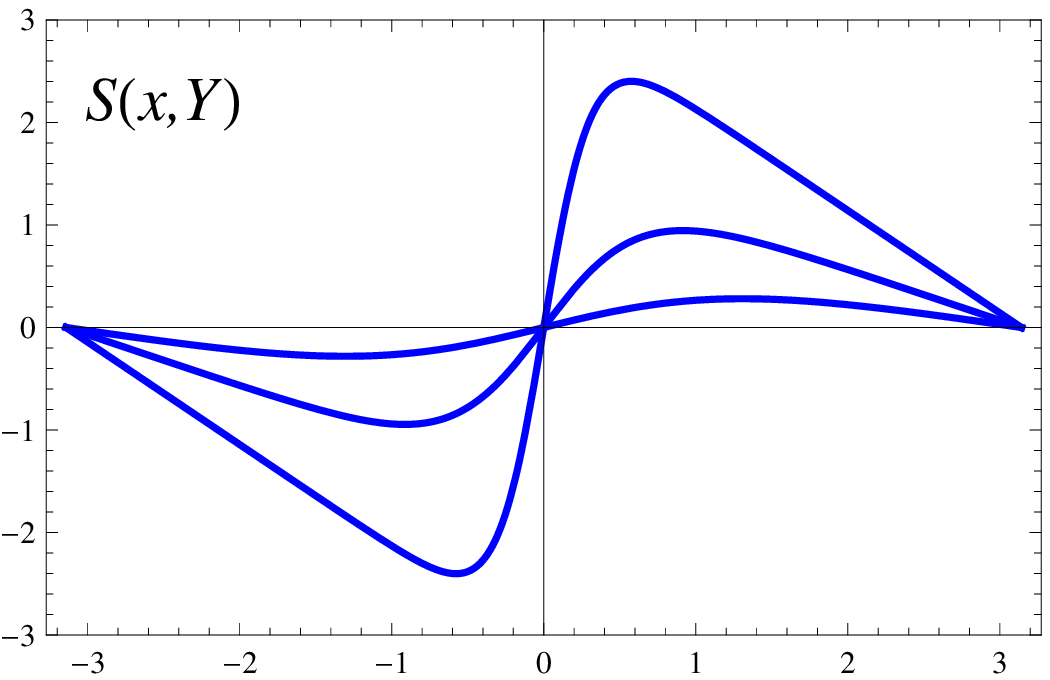,height=50mm,width=80mm}
\end{center}
\caption{Functions $C(x,Y)$ and $S(x,Y)$ for $Y=0.5$ (larger
  amplitude), $1$ and $2$ (smaller amplitude).}
\label{fig:CS}
\end{figure}
The solutions $Q_{1,2}(x)$ can also be expressed as a convolution (for
simplicity we take the symmetric case $Y_1=-Y_2=Y$)
\begin{equation}
Q_r(x) =  \frac{1}{2\pi}\tilde{Q}_{r,0} + \int\frac{d\phi}{2\pi} \Big(
C(\phi,Y) \, P_R(x-\phi)
 - \sigma_r   \,
S(\phi,Y) \, P_I(x-\phi)
\Big)
,
\quad r=1,2,
\qquad
\sigma_{1,2} \equiv \pm 1
.
\end{equation}
Here $P_{R,I}(x)$ stand for the real and imaginary parts of $P(x)$, and
\begin{equation}
C(x,Y) \equiv  \sum_{n=1}^\infty \frac{\cos(n x)}{\cosh(nY)}
,\qquad
S(x,Y) \equiv  \sum_{n=1}^\infty \frac{\sin(n x)}{\sinh(nY)}
.
\end{equation}
Clearly as $Y$ increases the functions $C$ and $S$ go uniformly to zero. This
ensures that eventually the conditions (\ref{eq:2.33a}) on $\hat{Q}_{1,2}(x)$
are met so that $Q_{1,2}(x)\ge 0$. The functions $C(x,Y)$ and $S(x,Y)$ are
displayed in Fig.~\ref{fig:CS}.

\begin{figure}[h]
\begin{center}
\epsfig{figure=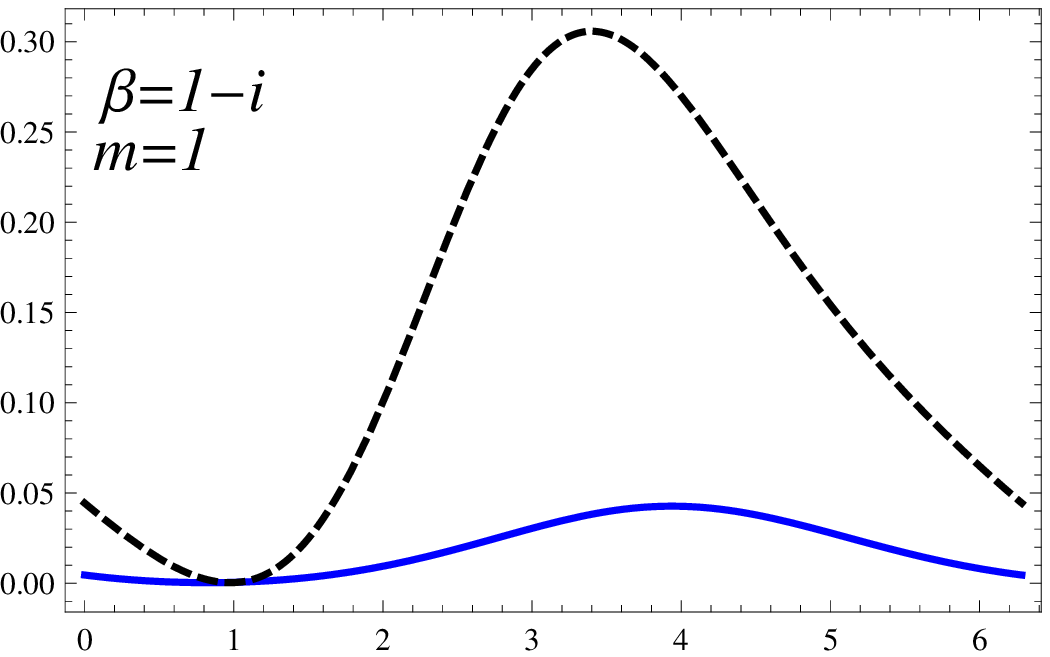,height=50mm,width=80mm}
\epsfig{figure=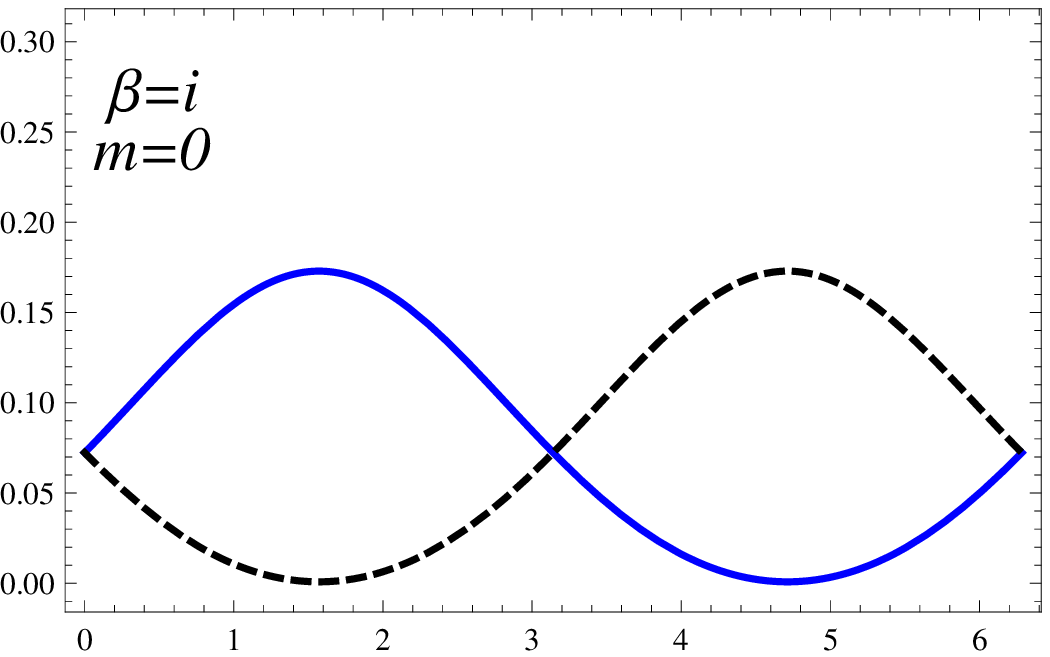,height=50mm,width=80mm}
\end{center}
\caption{Functions $Q_1(x)$ (solid line) and $Q_2(x)$ (dashed line) for the
  two-branches representation of the periodic action $S(x) = \beta
  \cos(x)+imx$. Left: $\beta=1-i$ and $m=1$, with $Y=1.17$. Right: $\beta=i$,
  $m=0$, with $Y=0.93$.}
\label{fig:2branchI}
\end{figure}
Examples of two-branches representations for the periodic case are shown in
Fig. \ref{fig:2branchI} for the action 
\begin{equation}
S(x)=\beta\cos(x)+imx,
\quad 
 \beta\in \C, ~ m\in\Z \,,
\end{equation}
using optimal values of $Y$.  The numerical solutions have been obtained from
the Fourier modes which are Bessel functions.

The case $m=0$ and $\beta$ imaginary is particularly interesting, since this
is one of the few cases (the other being a quadratic action, or a real action)
in which the equilibrium Fokker-Planck equation of the complex Langevin
algorithm can be solved analytically in $\rho(z)$. Unfortunately, that
solution turns out to be wrong, giving $\esp{e^{-ikz}}_{{\rm
    CL}}=\delta_{k,0}$, instead of the correct Fourier modes of $P$.

The existence of two-branches representations for very general complex
probabilities proves in particular that representations exist which are valid
for observables which are entire functions, no matter how wildly behaved they
are at infinity in the imaginary direction, provided they are convergent at
infinity in the real direction (in the non compact case). On the other hand,
representations that fill the complex plane would be problematic for those
wild observables; since the integral of $A(z)\rho(z)$ would not be absolutely
convergent the variance would diverge, rendering a Monte Carlo approach
useless.

Another observation is that the use of more than two branches does not seem to
introduce any improvement regarding localization of the support of the
representation.

\subsubsection{Representations in higher dimensions}
\label{sec:2.C.4}

The construction of representations by means of convolutions, either in the
compact or non compact cases, expresses that
\begin{equation}
Q_r = C_r* P + C_r^* * P^*
,\quad r=1,2
\end{equation}
involving two known universal functions $C_{1,2}(x,Y_1,Y_2)$. Unfortunately,
this simple scheme does not immediately extend to higher dimensions, although
representations based on the Fourier modes exist in the periodic case
\cite{Salcedo:2007ji}.  Counting degrees of freedom one can expect that
regardless the number of dimensions, a complex distribution can always be
traded by two real (and hopefully positive) distributions, however
unless the construction of $Q_{1,2}$ is local, that is, updating one coordinate
depends on only a few other coordinates, the method will not be useful in the
many dimensional case, that is, when a Monte Carlo approach is needed.

Nevertheless, it is of interest to discuss the construction of optimal
representations in higher dimensions. We consider the two-dimensional case, as
the ideas involved can be extrapolated to the general case. Also we take the
periodic case which is simpler.

Let $P(x_1,x_2)$ be a periodic two-dimensional normalized complex probability,
\begin{equation}
\iint_0^{2\pi} P(x_1,x_2) \,  dx_1 dx_2   = 1
.
\end{equation}

One way to proceed is by reducing the dimension. Let $P(x_2)$ be the marginal
distribution of $x_2$, and $P(x_1|x_2)$ the conditional probability of $x_1$
\begin{equation}
 P(x_2) = \int_0^{2\pi} P(x_1,x_2) \,  dx_1
,\qquad
P(x_1|x_2) = \frac{P(x_1,x_2)}{P(x_2)}
.
\end{equation}
Now, let $\rho(z_2)$ be a representation of $P(x_2)$ and $\rho(z_1|z_2)$ a
representation of $P(x_1|z_2)$ regarded as a function of $x_1$, and where
$P(x_1|z_2)$ refers to the analytical extension of $P(x_1|x_2)$. These are
regular one-dimensional distributions that we know how to sample. Then
$\rho(z_1,z_2) \equiv \rho(z_1|z_2)\rho(z_2)$ is a representation of
$P(x_1,x_2)$. Indeed, for any observable $A(x_1,x_2)$
\begin{equation}\begin{split}
\esp{A}_\rho &= 
\int A(z_1,z_2)\rho(z_1|z_2)\rho(z_2) \, d^2z_1 d^2z_2
\\
&=
\int A(x_1,z_2) P(x_1|z_2)\rho(z_2) \, dx_1 d^2z_2
=
\int A(x_1,x_2) P(x_1|x_2)P(x_2) \, dx_1 dx_2
=
\esp{A}_P
.
\end{split}\end{equation}

One drawback with the approach just presented is that the analytical extension
of $P(x_1,x_2)$ is needed, and also that $P(z_2)$ might have zeros on the
complex plane. These problems can be solved by means of the following
trick. Write $P$ as
\begin{equation}
P(x_1,x_2) = \frac{1}{2\pi}(P(x_2) - P^\prime(x_2) ) + P^\prime(x_1,x_2) 
,
\label{eq:2.40}
\end{equation}
where $P^\prime(x_2)$ is any positive distribution with normalization less
than one. In this case $P(x_2) - P^\prime(x_2)$ is one-dimensional,
normalizable and easily representable. Also, the marginal probability of the
remainder $P^\prime(x_1,x_2)$ is just $P^\prime(x_2)$. Since this marginal
probability is already positive one can apply the previous method to
$P^\prime(x_1,x_2)$ and a representation of $P^\prime(x_2)$ is not needed.
Regarding the choice of the auxiliary probability $P^\prime(x_2)$, we note that
a nice (i.e., localized) representation of $P^\prime(x_1|x_2)$ favors a
$P^\prime(x_2)$ as large as possible, while the representation of $P(x_2) -
P^\prime(x_2)$ favors a small $P^\prime(x_2)$, so a compromise has to be
taken.

Another approach mimics the two-branches method discussed for the
one-dimensional case. The natural proposal is (already taking a symmetric
choice, this is unessential)
\begin{equation}
P(\vec{x}) = \sum_{r=1,2} Q_r(\vec{x} -i \sigma_r \vec{Y} )
,
\label{eq:2.41}
\end{equation}
where $Q_r(\vec{x})$ are positive functions, $\vec{Y}=(Y_1,Y_2)$ and
$\sigma_r=\pm 1$ for $r=1,2$. Introducing Fourier modes
\begin{equation}
P(\vec{x}) = \frac{1}{(2\pi)^2} \sum_{\vec{k}} e^{i \vec{k}\cdot\vec{x}}
  \tilde{P}_{\vec{k}}
,
\end{equation}
one readily obtains the solution
\begin{equation}
\tilde{Q}_{r,\vec{k}} = \sigma_r \frac{
e^{\sigma_r \vec{k}\cdot\vec{Y}} \tilde{P}_{\vec{k}}
-e^{-\sigma_r \vec{k}\cdot\vec{Y}} \tilde{P}^*_{-\vec{k}}
}{2\sinh(2\vec{k}\cdot\vec{Y})}
.
\end{equation}
An obstruction arises here for the modes $\vec{k}\cdot\vec{Y}=0$. For them the
formula is only consistent if $\tilde{P}_{\vec{k}}=\tilde{P}^*_{-\vec{k}}$
and in this case $\tilde{Q}_{r,\vec{k}}=\tilde{P}_{\vec{k}}/2$.

\begin{figure}[h]
\begin{center}
\epsfig{figure=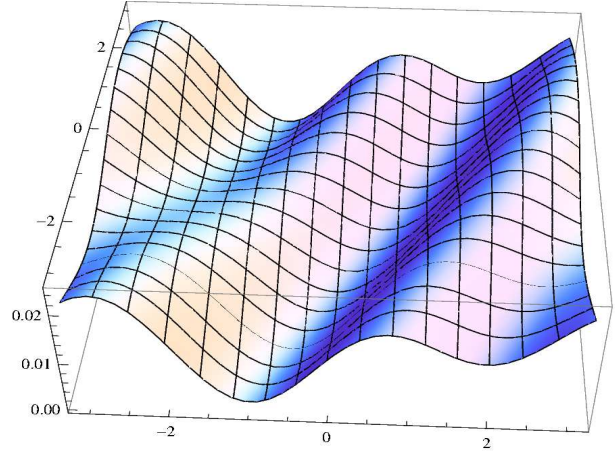,height=60mm,width=90mm}
\end{center}
\caption{Representation of $P \propto g(x_1)g(x_2)g(x_1-x_2)$ for $g(x)=1+2a
  \cos(x)$ with $a=i$, using $Y_2=3Y_1$. The function $Q_1(x_1,x_2)$ is
  displayed on $[-\pi,\pi]\times[-\pi,\pi]$ for $Y_2=4.43$. This is the
  optimal value, that is, $\min Q_1=0$.}
\label{fig:3D}
\end{figure}
There is no obstruction when $\vec{Y}$ is chosen so that $\vec{k}\cdot\vec{Y}
\not= 0$ unless $\vec{k}=\vec{0}$ (similarly to the one-dimensional case,
$\vec{k}=\vec{0}$ poses no problem due to $\tilde{P}_{\vec{0}}=1$). For
example, consider the complex distribution
\begin{equation}
P(\vec{x}) = N g(x_1)g(x_2)g(x_1-x_2)
,
\qquad
g(x) \equiv 1 + 2 a \cos(x)
,\quad a\in\C
.
\label{eq:2.44}
\end{equation}
The only relevant Fourier modes are $k_{1,2} = 0,\pm1,\pm2$, so the choice
$\vec{Y} = (Y,3Y)$ guarantees that $\vec{k}\cdot\vec{Y} = 0$ only for
$\vec{k}=\vec{0}$.  The complex probability proposed in \Eq{2.44} can be
represented with positive $Q_{1,2}(\vec{x})$ by taking a sufficiently large
value of $Y>0$. This is displayed in Fig. \ref{fig:3D} for $a=i$. There we
have taken $\tilde{Q}_{r,\vec{0}}=\frac{1}{2}$ for $r=1,2$, and automatically
$Q_2(\vec{x})=Q_1(-\vec{x})$. Similar results are obtained for
$Y_2/Y_1=\sqrt{2}$, which obviously also guarantees $\vec{k}\cdot\vec{Y} \not=
0$.  It is interesting that this probability distribution would be beyond a
complex Langevin approach, as $P(z_1,z_2)$ has zeros on $\C^2$.

When many Fourier modes are involved it is not possible to prevent small values
of $\vec{k}\cdot\vec{Y}$ and a more direct solution has to be adopted.
Without loss of generality, let us assume that our choice is $\vec{Y}= (Y,0)$
hence the problematic modes are those with $k_1=0$.  This choice implies that
only $x_1$ is moved to the complex plane,
\begin{equation}
P(x_1,x_2) = \sum_{r=1,2}Q_r(x_1 -i \sigma_r Y,x_2 )
.
\label{eq:2.41a}
\end{equation}
Clearly, this equation is only consistent when the marginal probability
$P(x_2)$ is positive. If this is not the case, the solution is to use the
trick described above, namely, to use \Eq{2.40} choosing a positive
$P^\prime(x_2)$. Now the two-dimensional version of the two-branches method
works for $P^\prime(x_1,x_2)$, and the one-dimensional version works for
$P(x_2) - P^\prime(x_2)$. Using this technique for the example in \Eq{2.44}
with $a=i$, a value $Y=2$ suffices for the two-dimensional representation and
$Y=2.45$ for the one-dimensional one. Variances are also smaller than using
$\vec{Y} = (Y,3Y)$.

\begin{figure}[h]
\begin{center}
\epsfig{figure=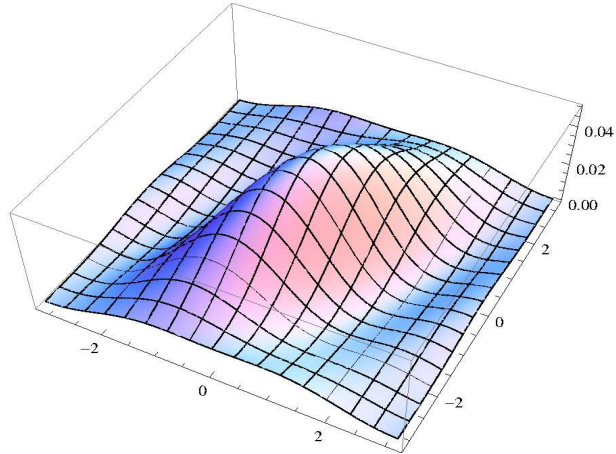,height=60mm,width=90mm}
\end{center}
\caption{Same as Fig. \ref{fig:3D} with $a=1$ and $Y_2=3Y_1=2.17$.}
\label{fig:3Db}
\end{figure}
As is well known, the real Langevin approach to sampling ordinary probability
distributions is affected by the segregation problem: the random walk cannot
cross the submanifold $P(x)=0$. When $P(x)$ is real but with negative and
positive regions, the same problem is inherited by the complex version of the
algorithm, the random walk gets trapped in one of the positive or negative
connected regions \cite{Fujimura:1993cq}. Such problem does not exist here.
For $a=1$, the probability distribution $P(x_1,x_2)$ of \Eq{2.44} is real and
changes sign whenever the argument of any of the factors $g(x)$ takes the
values $x=\pm\arccos(-1/2)$.  In Fig. \ref{fig:3Db} a representation of this
$P(x_1,x_2)$ is presented, using $Y_2=3Y_1 = 2.17$.

\section{Complex heat bath}
\label{sec:3}

As already noted the Monte Carlo method is required when the number of degrees
of freedom is large. Further, the sampling algorithm must be local, in the
sense defined in the Introduction, in order for the implementation not to be
prohibitively expensive. A local implementation can be done for particular
actions or Hamiltonians which admit some kind of short-cut allowing to turn
the complex probability problem into a standard one. Another local approach for
complex probabilities is that of reweighting, by sampling a local and positive
auxiliary probability distribution. Often $|P(x)|$ is used for this
purpose. As is well known, this approach works nicely for small systems but
becomes inefficient for larger ones where the expectation value of the phase of $P$ gets
close to zero \cite{Bhanot:1987nv}.  Among the approaches based on
constructing a representation of $P(x)$, only the complex Langevin algorithm
preserves locality.  This makes the algorithm very attractive. Unfortunately,
this method lacks a solid mathematical basis, and in fact it does not always
produce correct results \cite{Salcedo:1993tj,Aarts:2011ax}.

\subsection{Complex Gibbs sampling approach}
\label{sec:3.A}

These limitations make it worthwhile to explore alternative approaches. The
one we analyze here is to extend the standard heat bath approach to the case
of complex probabilities. In the standard Gibbs sampling the configuration
follows a Markovian walk and each variable is updated in turn using its
conditional probability, keeping the other variables fixed in that step. In
our complex version, the configuration follows a Markovian walk on the complex
manifold. Each variable is updated in turn using a {\em representation} of its
conditional probability, conditional to the values of the other variables,
which in general will be complex. This relies on the analytical extension of
the complex probability to the complex manifold, in the same way as in the
complex Langevin algorithm.

Let us remark that the need of an analytical extension of $P(x)$ is a
limitation of the complex Gibbs sampling approach. Representations exist for
any (or very general) complex probabilities. Having one such representation,
any standard sampling method could be used, although the issue of localization
on the complex manifold still persists. Unfortunately, such representations in
high dimensions are not easy to obtain through any useful (local)
approach. The virtue of a complex Gibbs sampling is that one does not need to
obtain representations of multidimensional complex distributions, since the
conditional probability depends just on the one variable to be updated.

Let us show that this procedure is formally correct. To make our point it is
sufficient to consider just two variables. Let $P(x_1,x_2)$ be the complex
probability in $\R^2$, and let $\rho(z_1,z_2)$ be a representation of it, a
positive distribution on $\C^2$.  We want to verify that the updated
distribution is also a representation. If the variable $z_1$ is updated to a
new value $z_1^\prime$, keeping $z_2$ fixed, the new distribution on $\C^2$
will be
\begin{equation}
\rho^\prime(z_1^\prime,z_2) = 
 \rho_{\rm rep}(z_1^\prime|z_2) \rho(z_2)
.
\label{eq:3.1}
\end{equation}
Here $\rho(z_2)$ is the marginal distribution of $z_2$ and $\rho_{\rm
  rep}(z_1^\prime|z_2)$ is a representation of the conditional probability
$P(x_1^\prime|z_2)$ (regarded as a function of $x_1^\prime$ only):
\begin{equation}
\rho(z_2) = \int d^2z_1\, \rho(z_1,z_2)
,\qquad
P(x_1^\prime|z_2) = \frac{P(x_1^\prime,z_2)}{P(z_2)}
.
\end{equation}
Likewise, $P(z_2) = \int dx_1\, P(x_1,z_2)$ where $P(z_1,z_2)$ refers to the
complex probability analytically extended to complex plane. $P(z_2)$ can also
be regarded as the analytical extension of the marginal probability $P(x_2)$.

In order to show that $\rho^\prime(z_1^\prime,z_2)$ is also a representation
of $P$ let us consider a generic (holomorphic) observable, $A(z_1,z_2)$, then
\begin{equation}\begin{split}
\int d^2z_1^\prime \, d^2z_2\, \rho^\prime(z_1^\prime,z_2) A(z_1^\prime,z_2) &=
\int d^2z_1^\prime \, d^2z_2\, \rho_{\rm rep}(z_1^\prime|z_2) \rho(z_2)
A(z_1^\prime,z_2)
\\ &=
\int dx_1^\prime \, d^2z_2\, P(x_1^\prime|z_2) \rho(z_2)
A(x_1^\prime,z_2)
\\ &=
\int dx_1^\prime \int d^2z_1\,d^2z_2\, \frac{P(x_1^\prime,z_2)}{P(z_2)} 
\rho(z_1,z_2) A(x_1^\prime,z_2)
\\ &=
\int dx_1^\prime \int dx_1\,dx_2\, \frac{P(x_1^\prime,x_2)}{P(x_2)} 
P(x_1,x_2) A(x_1^\prime,x_2)
\\ &=
\int dx_1^\prime \,dx_2\, P(x_1^\prime,x_2) A(x_1^\prime,x_2)
.
\end{split}\end{equation}
That is, $\esp{A}_{\rho^\prime}= \esp{A}_P$ and $\rho^\prime$ is a
representation of $P$. In the second equality it has been used that $\rho_{\rm
  rep}(z_1^\prime|z_2)$ is a representation of $P(x_1^\prime|z_2)$ with
respect to $z_1^\prime$. In the fourth equality it has been used that
$\rho(z_1,z_2)$ is a representation of $P(x_1,x_2)$ and the other factors
depend analytically on $z_1$ and $z_2$.
 
Note that \Eq{3.1} is not a standard heat bath sampling of $\rho(z_1,z_2)$:
there $z_1$ would be updated using $\rho(z_1^\prime|z_2)=
\rho(z_1^\prime,z_2)/\rho(z_2)$ instead of $\rho_{\rm rep}(z_1^\prime|z_2)$.
In the standard sampling $\rho(z_1,z_2)$ is unchanged under updates whereas
in the complex Gibbs sampling all one shows is that the new distribution is
still a representation, but not necessarily the same as before the update. 

Of course, a standard Gibbs sampling of $\rho(z)$ would be completely correct
(and in fact preferable), but the problem is that $\rho(z)$ is not available,
we only know how to construct representations of one- (or at any rate low-)
dimensional complex distributions such as $P(x_i|\{z_{j\not=i}\})$, and hope
that the Markovian chain converges to the correct result. We have shown that
the property of being a representation is preserved, however, the powerful
theorems that apply for Markovian chains of positive probability distributions
are not guaranteed to work in the complex case. In this sense, the complex
Gibbs sampling lacks a sound mathematical basis, as is also the case for other
approaches, like the complex Langevin algorithm. 

An specific way in which our complex heat bath sampling may find trouble is
related to the need of analytical extension: even if the marginal probability
$P(\{x_{j\not=i}\})$ is never zero on the real manifold, the actual algorithm
depends on the marginal probability on the complex manifold through analytical
extension, $P(\{z_{j\not=i}\})$. Zeroes in this function are potentially
problematic; proximity to a zero implies a highly non-positive definite
distribution and this requires going deeply into the complex plane. 
\ignore{ In this
regard, an improvement would be to update pairs (or more) of variables. It is
expected that representing two variables simultaneously allows to remain
closer to the real axis, as compared to representing first one variable and
then the other (this is obviously true if the total number of variables is
two). This entails the construction of representations in more than one
dimension, as discussed in Sec. \ref{sec:2.C.4}.
}

\subsection{Complex Gaussian action}
\label{sec:3.B}

Preliminary tests on simple distributions with few variables indicate that the
approach may work, however, the situation might be different for large
systems.  In order to test the complex Gibbs sampling proposal in
many-dimensional settings, we have studied a $d$-dimensional hypercubic
lattice of size $N$ in each direction. First we consider a quadratic action
with nearest-neighbor complex coupling.

The partition function is
\begin{equation}
Z(\beta) = \int e^{- S[\phi]}  \prod_x d\phi_x 
.
\end{equation}
It is defined through integration over the $V=N^d$ variables $\phi_x$ taking
real values, although during the Monte Carlo simulation the $\phi_x$ will
become complex in general. The action is given by
\begin{equation}
S[\phi] = \sum_x \left(
 \phi_x^2 + \beta \phi_x \sum_{\mu=1}^d \phi_{x+\hat{\mu}}
\right)
,
\qquad \beta\in \C
\,,
\label{eq:1a}
\end{equation}
and we adopt periodic boundary conditions. The action is complex by allowing
$\beta$ to be complex.

This case is simple enough to have analytic expressions of the expectation
values of typical observables. Also, representation of the $V$-dimensional
distribution $P(x)$ can be obtained in the present case. For this action the
implementation of the complex heat bath algorithm is straightforward. Indeed,
the conditional probability of the variable $\phi_x$ is
\begin{equation}
P(\phi_x|\{\phi_{x^\prime\not=x}\})
\propto
\exp(-\phi_x^2-\beta\phi_x\hat{\phi}_x)
\propto
\exp\left(-(\phi_x + \frac{\beta}{2}\hat{\phi}_x)^2 \right)
,
\qquad
\hat{\phi_x} \equiv \sum_{\mu=1}^d (\phi_{x+\hat{\mu}} + \phi_{x-\hat{\mu}} )
.
\end{equation}
Thus the update of the variable $\phi_x$ takes the simple form
\begin{equation}
\phi_x = \xi - \frac{\beta}{2}\hat{\phi}_x
,
\end{equation}
where the random variable $\xi$ is distributed according to $e^{-\xi^2}$.

\begin{table*}[h]
\label{tab:2}
\begin{tabular}{ >{$}l<{$}  >{$}l<{$} >{$}l<{$} >{$}l<{$} >{$}l<{$} >{$}l<{$}
 >{$}l<{$}     }
\Re\beta & 0.0 & 0.1 & 0.2 & 0.3 & 0.4 & 0.5  
\\
|\Im\beta| & 0.70 & 0.65 & 0.58 & 0.48 & 0.35 & 0.0
\end{tabular}\\
\caption{For a $2^4$ lattice, and for several values of $\Re\beta$, maximum
  values of $|\Im\beta|$ for which the complex bath algorithm converges, for
  the Gaussian action of \Eq{1a}.}
\end{table*}
We have checked through numerical experiments that whenever this Monte Carlo
simulation converges it does so to the correct expectation values. Clearly,
$\Im\beta$ does not affect $|P(\phi)|$, so the action is well-behaved, meaning
that the integrals involved are absolutely convergent, provided $\Re\beta$
stays within the appropriate limits (for even $N$ this is simply
$|\Re\beta|<1/d$) regardless of the value of $\Im\beta$. However, in our
complex version of the Gibbs sampling, convergence takes place only for
suitably bounded values of $\Im\beta$. This is shown in Table \ref{tab:2} for
a $2^4$ lattice.  The lack of convergence in otherwise well-posed problems
reflects the fact that the standard Markov Chain convergence theorems do not
immediately apply to the complex case, and in particular for our complex Gibbs
procedure.

\subsection{Complex $\lambda \phi^4$ action}
\label{sec:3.C}

In order to further check the approach we have considered a non-linear version
of the previous action, by adding a $\lambda\phi^4$ term. Specifically, we
study the action
\begin{equation}
S[\phi] = \sum_x \left(
\phi_x^4 + \phi_x^2 + \beta \phi_x \sum_{\mu=1}^d \phi_{x+\hat{\mu}}
\right)
,
\qquad \beta\in \C
\,,
\label{eq:1}
\end{equation}
again with hypercubic geometry $V=N^d$ and periodic boundary conditions.

The action enjoys some obvious symmetries (translations, lattice isotropy and
spatial reflection, as well as parity under $\phi_x \to -\phi_x$). In addition
when $N$ is an {\em even} number, a change $\beta \to -\beta$ can be
compensated by the transformation $\phi_x\to\pm\phi_x$, with plus/minus sign
for even/odd sites. Thus for even values of $N$, $Z(\beta)$ is an even
function of $\beta$. Also $Z(\beta)^* = Z(\beta^*)$, hence, when $N$ is even
$Z$ is real for purely imaginary $\beta$.

As observables to be estimated by Monte Carlo we take the following ones
\begin{equation}
O_1 = \frac{\beta}{V}\sum_x \phi_x \hat{\phi_x}
, \quad
O^\prime_1 = \frac{1}{V}\sum_x \phi_x \tilde{\phi}_x
,
\quad
O_2 = \frac{\beta}{2Vd}\sum_x \hat{\phi_x}^2
, \quad
O^\prime_2 = \frac{1}{2Vd}\sum_x \tilde{\phi}_x \hat{\phi_x}
,
\end{equation}
where the auxiliary fields are defined as
\begin{equation}
\hat{\phi_x} \equiv \sum_{\mu=1}^d (\phi_{x+\hat{\mu}} + \phi_{x-\hat{\mu}} )
,\qquad
\tilde{\phi}_x \equiv 4 \phi_x^3 + 2 \phi_x
.
\end{equation}
The observables $O_1$ and $O_2$ are quadratic while $O_1^\prime$ and
$O_2^\prime$ are quartic, and so they have larger variance than the former
ones.

\subsubsection{Strong coupling expansion}

In a strong coupling expansion in powers of $\beta$, $\log
Z(\beta)$ comes as a sum over closed paths on the lattice, formed by
consecutive links. Terms of order $\beta^n$ correspond to paths of length $n$.
For even $N$ all closed path have even length, so $Z(-\beta)=
Z(\beta)$. For odd $N$, contractile closed path have even length but
homotopically non trivial paths (winding through the periodic boundary
conditions) can have an odd length, hence introducing odd powers of $\beta$ in
$\log Z(\beta)$. These start at order $\beta^N$.

Within the strong coupling expansion, the expectation values of $O_1$ and
$O_2$ come as a series of powers of $\beta$ with real coefficients. The
leading order contributions are\footnote{This is for $N>2$. For $N=2$ the
  terms in \Eq{3.12} are of the same order and they have to be added to
  these ones.}
\begin{equation}
\esp{O_1} = -2d \beta^2 \esp{\phi^2}_0^2 + O(\beta^3)
,\qquad
\esp{O_2} = \beta \esp{\phi^2}_0 + O(\beta^2)
,
\qquad
\esp{\phi^2}_0 \equiv \esp{\phi^2}_{\beta=0} = 0.234
.
\end{equation}

Actually, $O_1$ relates sites with opposite parity, so for even (or infinite)
$N$ its expectation value has only even powers of $\beta$. In this case
$\esp{O_1}$ will be real for purely imaginary $\beta$. For odd $N$, odd powers
start at order $\beta^N$. These contributions come from non contractile closed
paths.\footnote{Of course, identical conclusion follows from the relation
  $\esp{O_1} = -2\beta/V \, \partial\log Z/\partial\beta$.} Likewise,
$\esp{O_2}$ has only odd powers of $\beta$ and it is imaginary for imaginary
$\beta$, except for odd $N$, in which case even powers start at
$O(\beta^{N-1})$. Inspection of $\esp{O_1^\prime}$ shows that in this regard,
this quantity behaves as $\esp{O_1}$, likewise $\esp{O_2^\prime}$ behaves as
$\esp{O_2}$. The leading contributions from topologically non trivial paths
are
\begin{equation}
\esp{O_1}_{\rm top} = -2d (-\beta)^N \esp{\phi^2}_0^N + O(\beta^{N+2})
,\qquad
\esp{O_2}_{\rm top} = -(-\beta)^{N-1} \esp{\phi^2}_0^{N-1} + O(\beta^{N+1})
.
\label{eq:3.12}
\end{equation}
These are the leading terms in $\Im\esp{O_1}$ and $\Re\esp{O_2}$ for imaginary
$\beta$ and odd $N$.

Since the non-linear action has no exact solution, we have monitored the
accuracy of the complex Gibbs sampling algorithm by using different checks. A
simple one is the fulfillment of the reality conditions on $\esp{O_1}$ and
$\esp{O_2}$ and as well as consistency with the strong coupling results for
small $\beta$.

\subsubsection{Transfer matrix}

Another check has been made by using a transfer matrix approach for $d=1$
(for $d>1$ the approach becomes prohibitive).  In this approach $Z =
\Tr(\hat{T}^N)$ where the transfer matrix $\hat{T}$ is an operator in
$L^2(\R)$ with kernel
\begin{equation}
\esp{\phi^\prime | \hat{T} | \phi}
=
e^{-\frac{1}{2}(\phi^\prime{}^4 + \phi^4 + \phi^\prime{}^2
+ \phi^2 + 2\beta \phi^\prime \phi )}
.
\end{equation}
The expectation value of observables of the type $A= F(\phi_{x+1},\phi_x)$,
such as $O_1$, $O_1^\prime$ and $O_2^\prime$,
can then be obtained as follows
\begin{equation}\begin{split}
\esp{A}
&=
\frac{\sum_n \lambda_n^N\esp{A}_n
}{\sum_n \lambda_n^N
}
,
\qquad
\esp{A}_n \equiv 
\lambda_n^{-1} \int d\phi^\prime d\phi
\bar{\psi}_n^*(\phi^\prime) \psi_n(\phi) 
 \esp{\phi^\prime | \hat{T} | \phi} F(\phi^\prime,\phi)
.
\end{split}\end{equation}
Here $\langle \bar{\psi}_n|$ and $|\psi_n\rangle$ are left and right
eigenvectors of $\hat{T}$ with eigenvalue $\lambda_n$ and normalized as
$\langle \bar{\psi}_n|\psi_m\rangle = \delta_{nm}$.  As it turns out, when
$\beta^2$ is a real number, $\hat{T}$ is a normal operator and it admits a
real and orthonormal eigen-basis. The eigenvalues are real for real $\beta$.
For imaginary $\beta$, half of the eigenvalues are real and the other half are
imaginary.

Using this approach,\footnote{To do this we have discretized $\phi$ along the
  same lines explained below for constructing representations of the
  conditional probability.} we find perfect agreement with the complex heat
bath results (whenever the latter method converges) for the cases studied,
$\beta= 0.5i$, $0.7i$ and $i$, with $N=20$. The complex heat bath algorithm
crashes for $\beta=2i$.

\subsubsection{Virial relations}

For higher dimensional lattices, a more useful check comes from the
(generalized) virial relations, also referred to as Schwinger-Dyson equations
or equations of motion. These exact relations state that for any (regular)
observable $A[\phi]$, and any site $x$,
\begin{equation}
\left\langle \frac{\partial A}{\partial \phi_x} \right\rangle
= \left\langle  A\frac{\partial S}{\partial \phi_x} \right\rangle
.
\label{eq:vir}
\end{equation}
The observables $O_1$, $O_2$, $O_1^\prime$ and $O_2^\prime$ arise naturally by
choosing $A=\phi_x$ and $A=\phi_{x+\hat{\mu}}$. For our action, the virial
relations imply that
\begin{equation}
\esp{ I_1} = \esp{I_2} = 0 
,\qquad
I_1 \equiv O_1 +
O^\prime_1 -1
,\qquad
I_2 \equiv O_2 + O^\prime_2
.
\end{equation}

\subsubsection{Complex Gibbs sampling implementation}

In addition, we compare the results obtained with the complex heat bath
approach with those obtained with standard reweighting. We sample
$e^{-\Re(S)}$ using a Metropolis algorithm and include the phase in the
observables. This technique is practical only for small lattices and/or small
$\Im\beta$ due to the overlap problem. Also, we compare with the complex
Langevin equation results since this has become a rather standard practice in
the context of complex probabilities.

In the three types of Monte Carlo calculations, reweighting (RW), complex
Langevin (CL), and complex heat bath (CHB), the expectation values and their
error are extracted from 20 independent runs with cold and hot starts. In RW
and CHB $10^5$ sweeps are applied, arranged in 100 batches of 1000 iterations
each, to monitor the thermalization. In CL each run has duration $10^4$ (in
Langevin time units), also
arranged in 100 batches of duration $10^2$ each. The CL step size is
controlled so that $\Delta t \le 0.001$ and $|\Delta\phi_x|\le 0.001$
\cite{Flower:1986hv,Aarts:2009dg}. In all three versions the 10 first batches
have been dropped as this was deemed sufficient to reach thermalization.

Since the CHB approach is new, some relevant details are in order. The
conditional probability needed in the complex Gibbs sampling takes the
following form
\begin{equation}
P(\phi_x|\{\phi_{x^\prime\not=x}\})
\propto
\exp(-\phi_x^4 - \phi_x^2-\beta\phi_x\hat{\phi}_x)
.
\end{equation}
For complex $\beta$ this distribution cannot be sampled directly and one has
to resort to representations with two branches, as explained in Sec.
\ref{sec:II.C.3}. In order to study the performance of the method in this
exploratory work, we have given priority to keeping things simple and under
control, leaving improvements in technical details of the implementation to
future work. The actual representation has been obtained by making use of
\Eq{2.25} and relying on a fast Fourier transform algorithm, applied back and
forth, to carry out the convolutions. To do this, $\Re\phi$ is restricted to a
box $[-L/2,L/2]$ and discretized with a step of size $h=L/2^K$. Typical values
are $L=20$, $40$, and $K=8$, $10$, $12$.  We have checked, by comparing with a
standard Metropolis calculation, that for real $\beta$ no sizable error is
introduced due to the finiteness of $L$ and $K$.

A representation, i.e., a pair $Q_{1,2}(x)$, is obtained for each site to be
updated. The minimal value of the parameter $Y$ (see \Eq{2.29}) is determined
by taking steps $\Delta Y$ starting at $|\Im\langle x\rangle| + \Delta Y$. For
$\Delta Y =0.1$, this starting value is already sufficient in more than
$99.99\%$ of the cases. In average, $1.4$ more steps are needed if $\Delta
Y=0.01$, with no noticeable gain in the accuracy of the expectation values.

During the Monte Carlo simulation one finds conditional probabilities
(controlled by the value of the environmental parameter $\hat{\phi}_x$ there)
with wildly different degrees of difficulty regarding its representation. For
$|\Im\beta|\le 0.5$ the soft cases are overwhelmingly predominant and for these
$Y$ remains small. The hard cases are rare but unavoidable, since the
parameter $\hat{\phi}_x$ can approach a zero of the normalization of the
conditional probability. The closest zero is at $\hat{\phi}_x = 4.60i$. For
these hard cases, $Y$ can attain huge values which could spoil the simulation
or make it crash. To prevent this we introduce a control parameter $Y_s$ such
that the site is not updated when a $Y$ larger that $Y_s$ would be
required. Since the decision of skipping the site (on that sweep only) is
taken ex post facto, some bias is introduced in the simulation.

For a too small value of $Y_s$ (i.e., a hard cutoff) the bias shows as a
violation of the virial relations $\langle I_{1,2}\rangle=0$. For too large
values, the variances increase and moreover the simulation may abort. For
$\beta=0.25i$ the value of $Y_s$ is not crucial. For $\beta=0.5i$, values of
$Y_s$ ranging from $2$ to $20$ allow to fulfill the virial relations with a
controlled noise. For $\beta=0.7i$, hard events become too frequent, making the
whole approach inviable: a too small $Y_s$ would be required, hence
introducing an unacceptable bias in the results, at least for the observables
$I_1$ and $I_2$. The situation resembles that of an asymptotic expansion,
e.g., $\sum_{n=0}^\infty (-1)^n n!  x^n$: the series must be truncated because
at some point the oscillatory terms start to grow. For small enough $x$ (in
our case $\beta$) many terms can be included and a good accuracy can be
attained, but the exact unbiased result is never obtained, unless some
resummation technique is applied.

Besides $Y_s$, a further regulator must be introduced. In principle the
optimal $Y$ is such that $\min_x \{Q_1(x), Q_2(x)\}=0$. In the soft cases this
poses no problem, however, for hard events such a strict requirement would
result in too large values of $Y$. In fact, for moderate values of $Y$ the
functions $Q_{1,2}(x)$ are positive where they are sizable, but may present
small negative tails in the region of large $x$. Removal of these small tails
is what forces $Y$ to be large. To address this problem we relax the
positivity requirement to $\min_x \{Q_1(x), Q_2(x)\} \ge -\epsilon$.  Values
$\epsilon \le 10^{-8}$ are too small, but $10^{-7}$ or $10^{-6}$ already yield
good results. In fact, even the larger value we have tried,
$\epsilon=10^{-3}$, turned out to be equally acceptable.

\subsubsection{Monte Carlo estimates}


\begin{table*}[t]
\caption{Expectation values (scaled by $10^3$) of $O_{1,2}$ and $I_{1,2}$ for
  several settings.  RW, CL and CHB refer to reweighting, complex Langevin and
  complex heat bath methods, respectively.  Standard deviations of the means
  are indicated in parenthesis; they affect the last digits shown. Each value
  is extracted from $20$ runs with hot and cold starts and $10^5$ sweeps
  ($10^4$ time units for CL) of which the last 90000 are used in the averages.
  Default parameters for CHB are $Y_s=5$, $\Delta Y=0.1$, and
  $\epsilon=10^{-6}$.  The mark $*$ indicates that $Y_s=\infty$ there. The
  mark $**$ indicates that data lying beyond $8$ standard deviations from the
  mean have been removed (see text).}
\label{tab:1}

\begin{tabular}{ >{$}l<{$} >{$}c<{$} >{$}r<{$} >{$}r<{$} >{$}l<{$} >{$}l<{$}
    >{$}l<{$} >{$}l<{$} >{$}l<{$} >{$}l<{$} >{$}l<{$} >{$}l<{$} l }

\beta & N^d & L & K 
& \multicolumn{2}{c}{$ 10^3 \times \langle O_1 \rangle $}  
& \multicolumn{2}{c}{$ 10^3 \times \langle O_2 \rangle $} 
& \multicolumn{2}{c}{$ 10^3 \times \langle I_1 \rangle $}  
& \multicolumn{2}{c}{$ 10^3 \times \langle I_2 \rangle $} 
& Method
\\ 
\hline
\ignore{
0.25i & 3^3 & 20 & 10
 & 19.751 \fz (22) &  - i\, 1.060 \fz (26)
 & \fm 3.000(16) &  + i\, 56.002(18)
 & -0.35 \fz (30) &  + i\, 0.17 \fz (19)
 & \fm 0.00 \fz (9) &  + i\, 0.04 \fz (7)
 & RW
\\
}

0.25i & 3^3 &  & 
 & 19.783 \fz (28) &  - i\, 1.134 \fz (33)
 & \fm 2.984 (12) &  + i\, 56.017 (19)
 & \fm 0.49 \fz (33) &  - i\, 0.07 \fz \fz (9)
 &   - 0.22 (10) &  - i\, 0.05 \fz (8)
 & RW
\\

\ignore{
0.25i & 3^3 &  & 
 & 19.693 \fz (18) &  - i\, 1.044 \fz (88)
 & \fm 2.989(15) &  + i\, 56.187(44)
 & \fm 5.03 \fz (73) &  - i\, 0.01 \fz\fz (9)
 & \fm 0.15  (26) &  + i\, 0.24 \fz (5)
 & CL
\\
}

0.25i & 3^3 &  & 
 & 19.740 \fz (21) &  - i\, 1.120 \fz (99)
 & \fm 3.009(17) &  + i\, 55.978(51)
 & \fm 0.97 \fz (73) &  - i\, 0.09 \fz\fz (8)
 & - 0.10  (30) &  + i\, 0.02 \fz (4)
 & CL
\\

0.25i & 3^3 & 20 & 10
 & 19.745 \fz (14) &  - i\, 1.099 \fz (28)
 & \fm 2.996(11) &  + i\, 55.967(16)
 & \fm 3.39(115) &  + i\, 0.09 \fz (23)
 & -0.12 \fz (9) &  - i\, 0.14 \fz (9)
 & CHB
\\
\hline

\ignore{
0.25i & 8^3 & 20 & 10
 & 19.730(173) &  + i\, 0.162(134)
 & \fm 0.072(49) &  + i\, 56.005(50)
 & \fm 0.63 \fz (55) &  - i\, 0.32 \fz (51)
 & \fm 0.49(41) &  + i\, 0.07(51)
 & RW
\\
}

0.25i & 8^3 &  &  
 & 19.985 (137) &  - i\, 0.153 (179)
 & - 0.074 (47) &  + i\, 55.958 (44)
 & \fm 0.59 \fz (63) &  + i\, 1.16 \fz (68)
 & - 0.55 (53) &  - i\, 0.67 (42)
 & RW
\\

\ignore{
0.25i & 8^3 &  & 
 & 19.723 \fz\fz (7) &  - i\, 0.016 \fz(18)
 & \fm 0.000 \fz (4) &  + i\, 56.055  (17)
 & \fm 1.81 \fz (12) &  + i\, 0.02 \fz\fz (2)
 & -0.02 \fz (6) &  + i\, 0.11 \fz (2)
 & CL
\\
}

0.25i & 8^3 &  & 
 & 19.746 \fz\fz (5) &  - i\, 0.021 \fz(20)
 & \fm 0.004 \fz (4) &  + i\, 55.977  (12)
 & \fm 0.29 \fz (24) &  - i\, 0.02 \fz\fz (2)
 & -0.03 \fz (7) &  + i\, 0.01 \fz (1)
 & CL
\\

0.25i & 8^3 & 20 & 10
 & 19.749 \fz\fz (4) &  + i\, 0.001 \fz\fz (8)
 & \fm 0.000 \fz (2) &  + i\, 55.969 \fz (5)
 & \fm 4.04 \fz (28) &  + i\, 0.04 \fz\fz (5)
 & -0.02 \fz (3) &  - i\, 0.13 \fz (2)
 & CHB
\\
\hline

\ignore{
0.5i & 3^3 & 20 & 10
 & 71.571(104) &  - i\, 6.301 \fz (95)
 & \fm 8.489(55) &  + i\, 99.889(32)
 & \fm 0.32 \fz (34) &  + i\, 0.19 \fz (31)
 & -0.13(15) &  - i\, 0.04(16)
 & RW
\\}

0.5i & 3^3 &  & 
 & 71.642(116) &  - i\, 6.146 \fz (74)
 & \fm 8.487(49) &  + i\, 99.851(47)
 & - 0.39 \fz (40) &  + i\, 0.53 \fz (28)
 & \fm 0.08 (14) &  - i\, 0.15 (21)
 & RW
\\

\ignore{
0.5i & 3^3 &  & 
 & 71.339 \fz (48) &  - i\, 5.984 (183)
 & \fm 8.492(62) &  + i\, 99.994(54)
 & \fm 4.08 (101) &  + i\, 0.01 \fz (13)
 & \fm 0.22(32) &  + i\, 0.17 \fz (5)
 & CL
\\
}

0.5i & 3^3 &  & 
 & 71.628 \fz (76) &  - i\, 5.985 (127)
 & \fm 8.567(49) &  \!\!\!\!+ i\, 100.073(98)
 & \fm 0.16 \fz (75) &  - i\, 0.07 \fz (15)
 & \fm 0.42(22) &  + i\, 0.03 \fz (9)
 & CL
\\

0.5i & 3^3 & 20 & 10
 & 71.578 \fz (41) &  - i\, 6.173 \fz (49)
 & \fm 8.510(32) &  + i\, 99.882(32)
 & \fm 4.70(125) &  + i\, 2.78(135)
 & \fm 0.45(22) &  + i\, 0.21(29)
 & CHB
\\
\hline

\ignore{
0.5i & 8^3 & 20 & 10
 &  9.3  (97) &   + i\, 19.(14)
 & \fm 7.8(60) &  + i\, 112.9(22)
 & \fm 53.(45) &  + i\, 05.(49)
 & \fm 31.(22) &  + i\, 93.(19)
 & RW
\\
}

0.5i & 8^3 &  & 
 &  -3.0  (37) &   + i\, 2.2(34)
 & \fm 1.8(38) &  + i\, 117.1(22)
 & - 03.(18) & - i\, 01.(17)
 & \fm 1.1(76) &  + i\, 118.(6)
 & RW
\\

\ignore{
0.5i & 8^3 &  & 
 & 71.197 \fz (14) &  + i\, 0.009 \fz (38)
 & - 0.013  (12) &  + i\, 99.764 (18)
 & \fm 2.07 \fz (16) &  + i\, 0.01 \fz\fz (3)
 & \fm 0.00 \fz (6) &  + i\, 0.27 \fz (3)
 & CL
\\
}

0.5i & 8^3 &  & 
 & 71.332 \fz (17) &  + i\, 0.076 \fz (32)
 & - 0.019  (18) &  + i\, 99.579 (18)
 & \fm 0.24 \fz (18) &  + i\, 0.06 \fz\fz (4)
 & \fm 0.08 \fz (6) &  - i\, 0.01 \fz (3)
 & CL
\\

0.5i & 8^3 & 20 & 10
 & 71.330 \fz (13) &  - i\, 0.003 \fz (13)
 & \fm 0.006 \fz (7) &  + i\, 99.568(10)
 & \fm 4.59 \fz (29) &  - i\, 0.01 \fz (20)
 & -0.01 \fz (5) &  - i\, 0.13 \fz (5)
 & CHB
\\

0.5i & 8^3 & 20 & 10
 & 71.352 \fz (11) &  + i\, 0.006 \fz (13)
 & -0.008 \fz (5) &  + i\, 99.570 \fz (8)
 & \fm 6.95(212) &  - i\, 1.22(264)
 & -0.06(12) &  - i\, 0.16(13)
 & CHB*
\\

0.5i & 8^3 & 20 & 8
 & 71.308 \fz (10) &  + i\, 0.013 \fz (12)
 & \fm 0.002 \fz (7) &  + i\, 99.571 \fz (7)
 & \fm 4.94 \fz (37) &  - i\, 0.11 \fz (20)
 & \fm 0.04 \fz (5) &  - i\, 0.07 \fz (5)
 & CHB
\\

0.5i & 8^3 & 40 & 8
 & 71.337 \fz (13) &  + i\, 0.011 \fz (14)
 & -0.001 \fz (7) &  + i\, 99.572 \fz (7)
 & \fm 1.40 \fz (66) &  + i\, 0.08 \fz (16)
 & \fm 0.04 \fz (6) &  - i\, 0.03 \fz (4)
 & CHB
\\

0.5i & 8^3 & 40 & 8
 & 71.337 \fz (13) &  + i\, 0.011 \fz (14)
 & \fm 0.000 \fz (7) &  + i\, 99.571 \fz (7)
 & \fm 0.10 \fz (14) &  + i\, 0.03 \fz (15)
 & \fm 0.02 \fz (4) &  - i\, 0.01 \fz (2)
 & CHB**
\\

0.5i & 8^3 & 40 & 12
 & 71.305 \fz\fz (8) &  - i\, 0.014 \fz (15)
 & \fm 0.004 \fz (9) &  + i\, 99.545 \fz (9)
 & \fm 2.44 \fz (85) &  + i\, 0.17 \fz (19)
 & \fm 0.00 \fz (5) &  - i\, 0.01 \fz (5)
 & CHB
\\

0.5i & 8^3 & 40 & 12
 & 71.305 \fz\fz (8) &  - i\, 0.014 \fz (15)
 & \fm 0.005 \fz (9) &  + i\, 99.544 \fz (9)
 & \fm 0.14 \fz (20) &  + i\, 0.02 \fz (13)
 & \fm 0.00 \fz (3) &  + i\, 0.03 \fz (3)
 & CHB**
\\
\hline

\ignore{
0.5i &16^3 &  & 
 & 71.249 \fz\fz (7) &  - i\, 0.023 \fz (13)
 & - 0.010 \fz (5) &  + i\, 99.709 \fz (8)
 & \fm 1.67 \fz\fz (8) &  + i\, 0.01 \fz\fz (2)
 & -0.04 \fz (3) &  + i\, 0.17 \fz (1)
 & CL
\\
}

0.5i &16^3 &  & 
 & 71.303 \fz\fz (7) &  - i\, 0.014 \fz (18)
 & - 0.016 \fz (7) &  + i\, 99.569 \fz (7)
 & \fm 0.11 \fz\fz (7) &  + i\, 0.02 \fz\fz (2)
 & -0.03 \fz (3) &  + i\, 0.03 \fz (1)
 & CL
\\

0.5i &16^3 & 20 & 10
 & 71.328 \fz\fz (5) &  - i\, 0.005 \fz\fz (6)
 & \fm 0.000 \fz (3) &  + i\, 99.567 \fz (4)
 & \fm 5.15 \fz\fz (9) &  - i\, 0.09 \fz\fz (7)
 & -0.01 \fz (2) &  - i\, 0.19 \fz (3)
 & CHB
\\
\hline

\ignore{
0.5i & 8^4 &  & 
 & 90.584 \fz\fz (7) &  + i\, 0.010 \fz (14)
 & - 0.003 \fz (7) &  + i\, 94.498 \fz (7)
 & \fm 1.41 \fz\fz (8) &  + i\, 0.02 \fz\fz (2)
 & \fm 0.01 \fz (2) &  + i\, 0.16 \fz (1)
 & CL
\\
}

0.5i & 8^4 &  & 
 & 90.652 \fz\fz (9) &  - i\, 0.013 \fz (12)
 & \fm 0.000 \fz (4) &  + i\, 94.365 \fz (9)
 & \fm 0.06 \fz\fz (9) &  - i\, 0.02 \fz\fz (3)
 & \fm 0.00 \fz (2) &  + i\, 0.03 \fz (1)
 & CL
\\

0.5i & 8^4 & 20 & 10
 & 90.670 \fz\fz (5) &  + i\, 0.005 \fz\fz (8)
 & \fm 0.000 \fz (3) &  + i\, 94.369 \fz (5)
 & \fm 3.19 \fz (20) &  - i\, 0.16 \fz (16)
 & \fm 0.02 \fz (2) &  + i\, 0.14 \fz (3)
 & CHB
\\
\hline

\end{tabular}\\
\end{table*}

\begin{figure}[t]
\begin{center}
\epsfig{figure=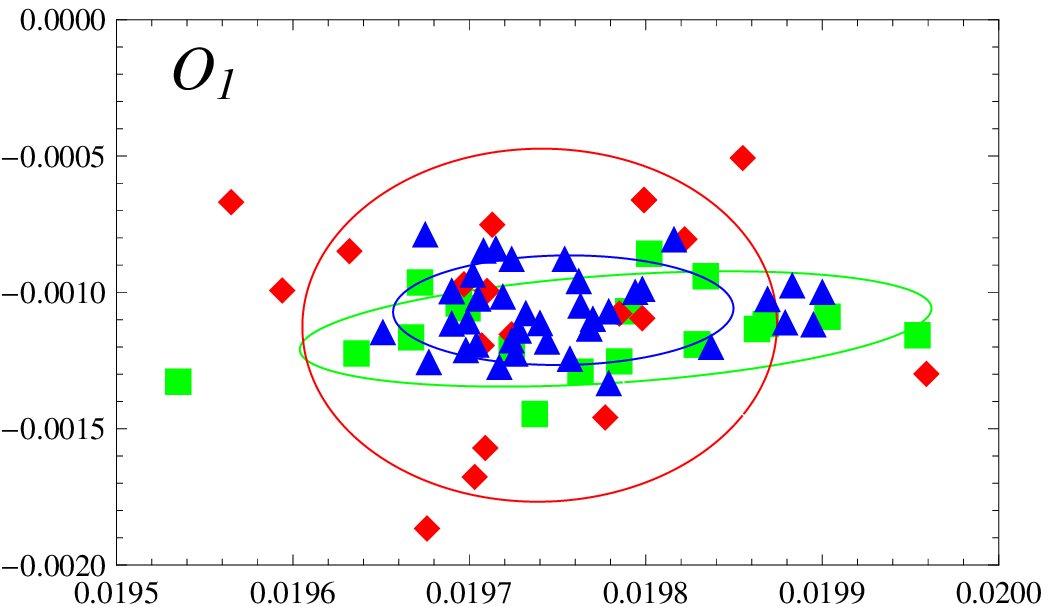,height=50mm,width=80mm}
\epsfig{figure=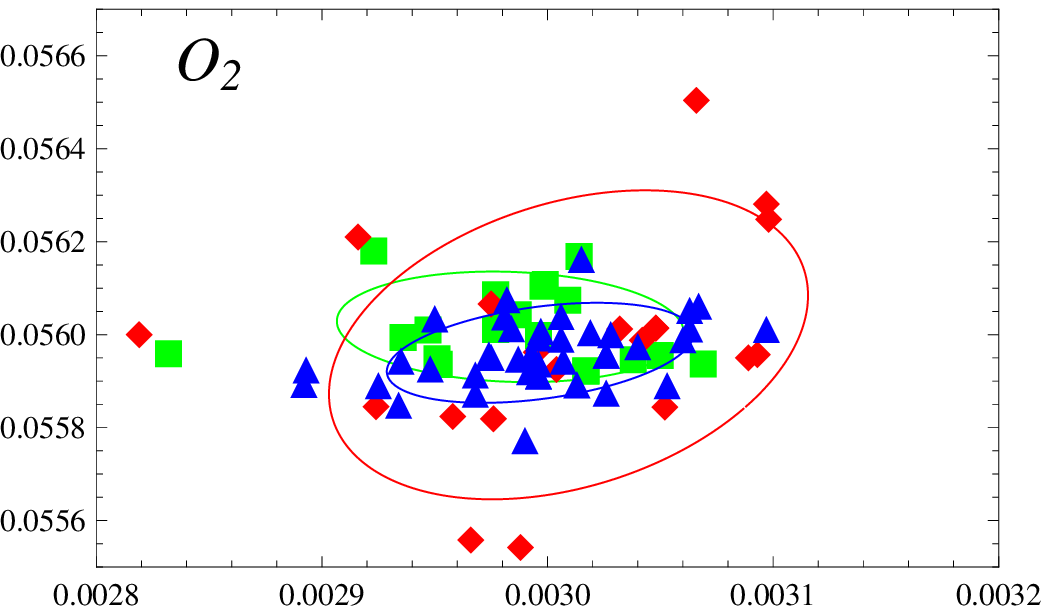,height=50mm,width=80mm}\\
\epsfig{figure=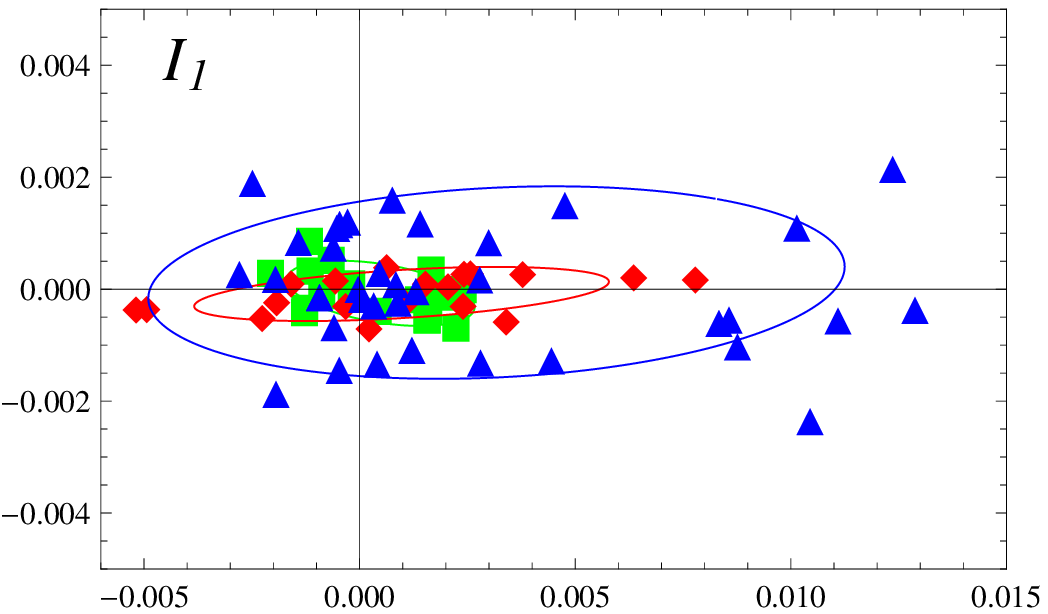,height=50mm,width=80mm}
\epsfig{figure=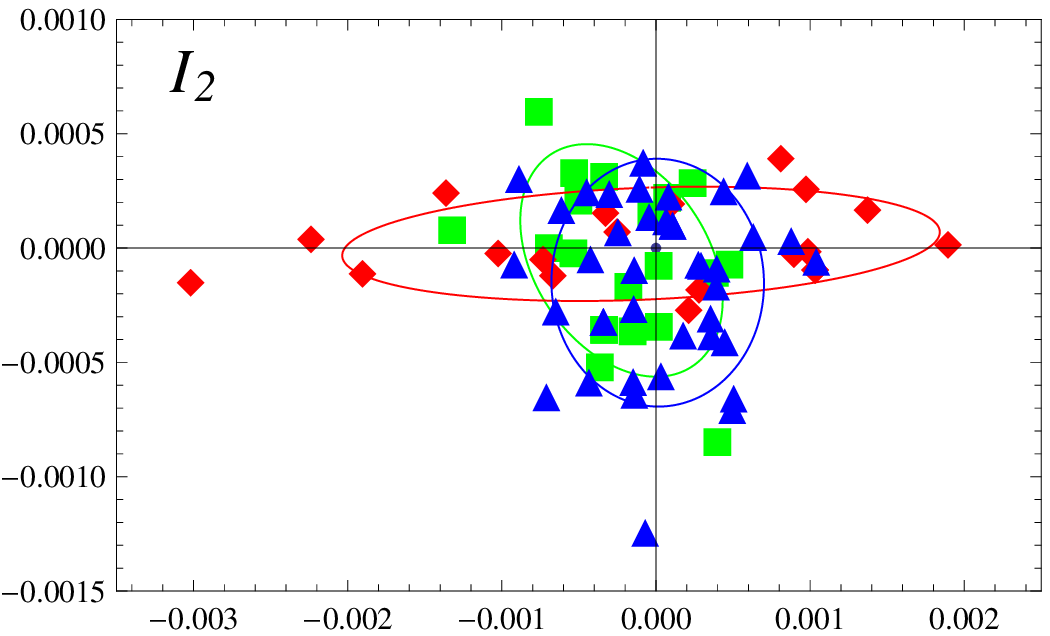,height=50mm,width=80mm}
\end{center}
\caption{Monte Carlo estimates of $O_1$, $O_2$, $I_1$ and $I_2$ for a $3^3$
  lattice with $\beta=0.25i$, from RW (green squares), CL (red rhombuses) and
  CHB (blue triangles). Each point represents one of the 20 independent
  runs. To guide the eye ellipses are extracted from mean values and variance
  matrices of each cloud. They are scaled to enclose $68\%$ of the Gaussian
  probability.}
\label{fig:chb4}
\end{figure}
Monte Carlo results for $\langle O_{1,2} \rangle$ and $\langle I_{1,2}
\rangle$ for purely imaginary $\beta$, obtained through our complex heat bath
(CHB) method, are displayed in Table \ref{tab:1}. Reweighting (RW) and complex
Langevin (CL) results are shown for comparison. One can see that CHB and CL
produce consistent results (with exception of $\Re\esp{I_1}$). RW also concurs
on the same values, when our calculation using this method can be trusted
(three first RW rows in Table \ref{tab:1}). 

As illustration, results for a $3^3$ lattice with coupling $\beta=0.25i$ are
shown in Fig. \ref{fig:chb4}. For the four observables, the points represent
the estimate obtained for each of the 20 independent Monte Carlo runs, for
each of the three versions. To guide the eye, we have analyzed the cloud of
points assuming a Gaussian probability distribution. The ellipses are scaled
to enclose $68\%$ of that Gaussian probability.

\begin{figure}[t]
\begin{center}
\epsfig{figure=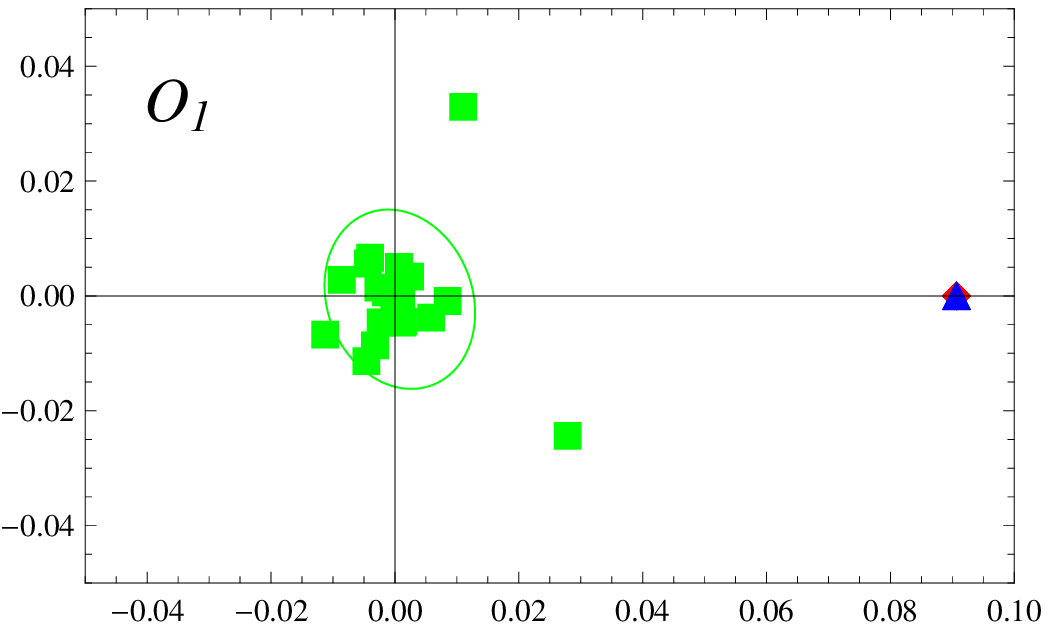,height=50mm,width=80mm}
\epsfig{figure=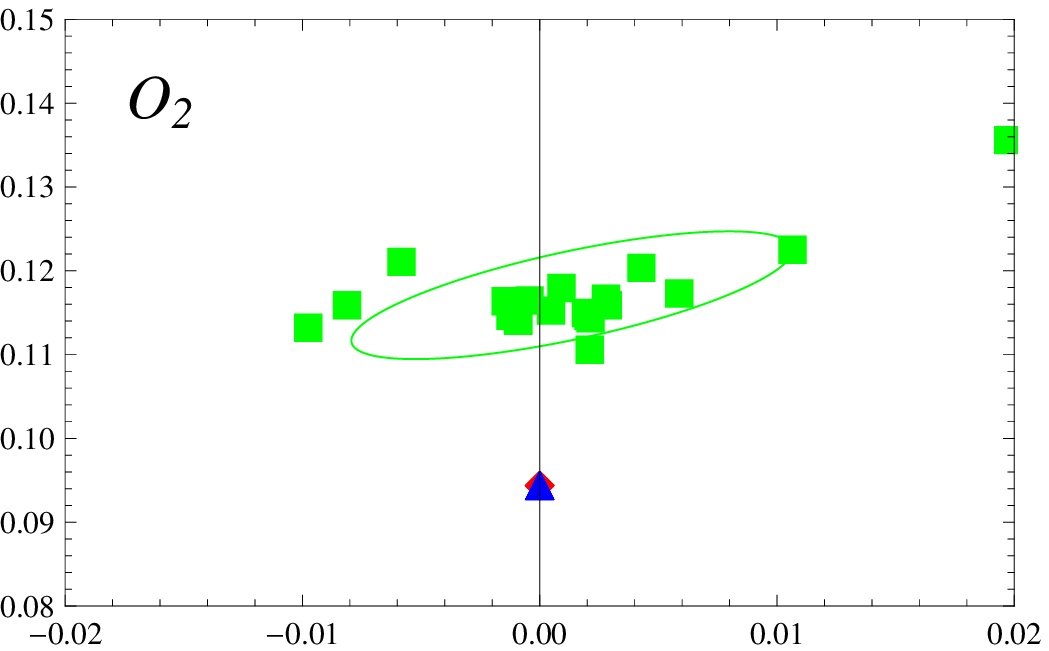,height=50mm,width=80mm}\\
\epsfig{figure=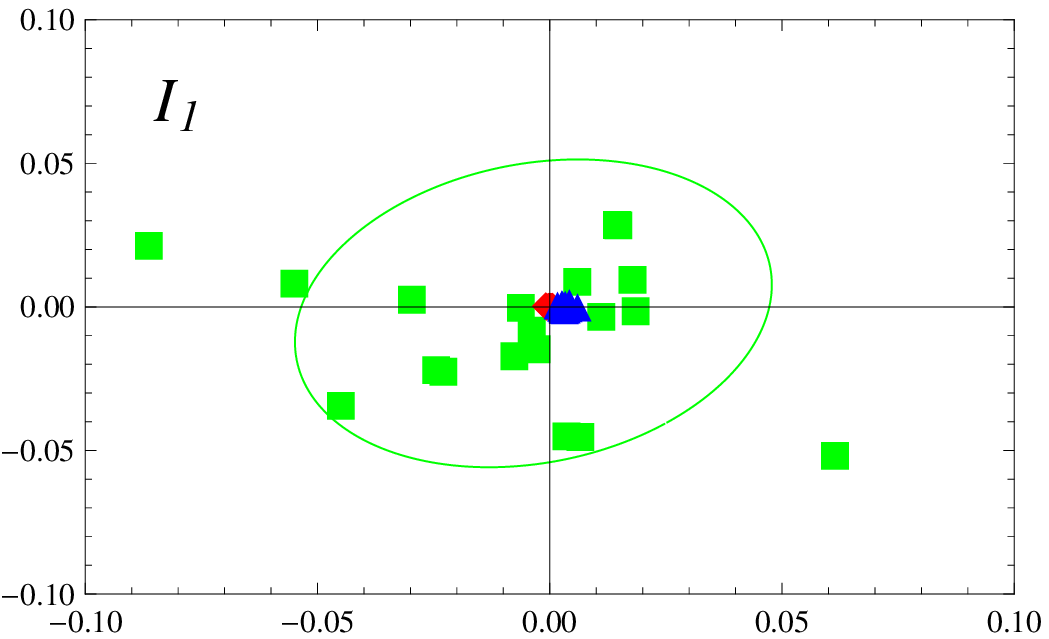,height=50mm,width=80mm}
\epsfig{figure=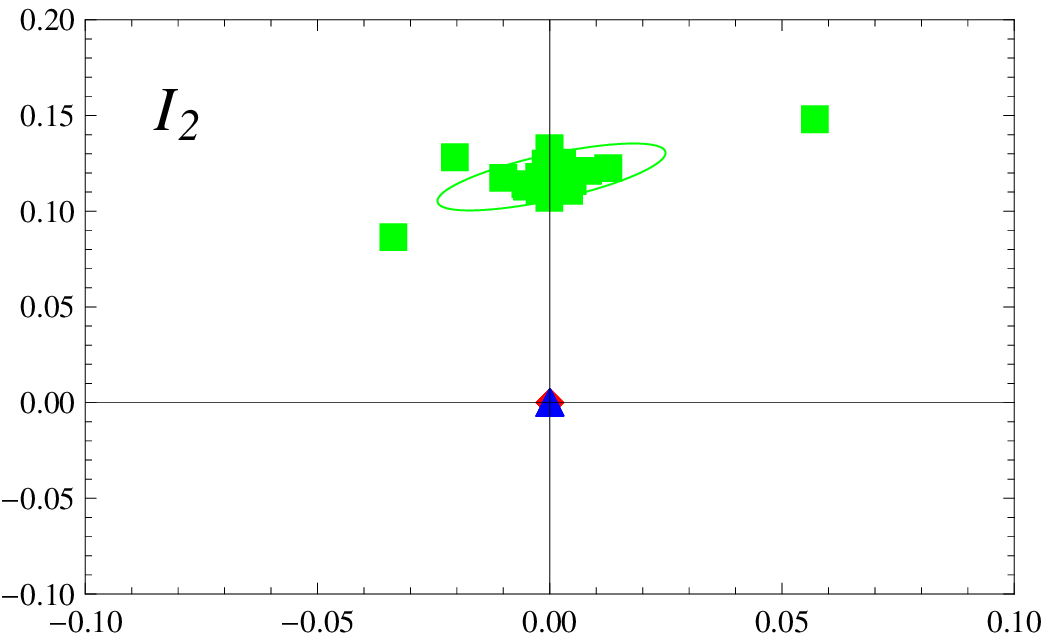,height=50mm,width=80mm}
\end{center}
\caption{Metastability of the RW calculation. Monte Carlo estimates of $O_1$,
  $O_2$, $I_1$ and $I_2$ for a $8^4$ lattice with $\beta=0.5i$, from RW (green
  squares), CL (red rhombuses) and CHB (blue triangles). For this lattice and
  coupling the RW estimate displays a net deviation (as well as enhanced
  dispersion). The incorrect RW estimates displayed actually reproduce the
  expectation values of the auxiliary $\Re S[\phi]$ (see text). Correct RW
  results would follow from using a sufficiently large number of sweeps.}
\label{fig:chb2}
\end{figure}

For $\beta=0.5i$ and $8^3$, the RW estimates presented in Table \ref{tab:1}
are not only noisy but also wrong, most clearly for $\Im\langle I_2\rangle$
which ought to be zero. By incorrect we mean that the cloud of points are well
separated from the correct results (see Fig. \ref{fig:chb2}) and so the
deviations cannot be merely attributed to the dispersion of the
points. Remarkably, this implies that consistent (yet incorrect) results are
obtained starting from quite different initial conditions (cold or hot) a
condition which is often invoked as a check of convergence of the Markovian
chain. That the convergence is actually metastable is signaled however by the
lack of fulfillment of the virial relations. Since such relations, \Eq{vir},
are always available for continuous degrees of freedom, they prove to be a
rather useful (necessary although not sufficient) test to check convergence.

We emphasize that the RW method itself has no bias, and the trouble comes
entirely from using a too short Markovian chain (in our case $10^5$
sweeps). If a sufficiently large number of sweeps were used, RW would yield
correct estimates with some noise. The latter can only be reduced by using
even more sweeps. Essentially\footnote{The discussion that follows is more
  literally suited to the case where the target distribution $P(x)$ is
  positive. When the weight is a phase, i.e., $|P/P_0|=1$, one has to rely on
  cancellation of phases (i.e., destructive interference) where they change
  rapidly or non cancellation (constructive interference) where they change
  slowly, as in the stationary phase method.} the problem is that in RW, the
sampling points (i.e., field configurations in our case) follow an auxiliary
distribution $P_0(x)$ instead of the target distribution $P(x)$, and the
weight proportional to $P/P_0$ is included in the observables:
\begin{equation}
\esp{A}_P  = \frac{\esp{wA}_{P_0}}{\esp{w}_{P_0}}
,\qquad
w(x) \equiv \frac{P(x)}{P_0(x)}
.
\end{equation}
In a proper implementation of RW, with sufficient (independent) sampling
points, most points lie where $P_0$ is important, not $P$.  The very few
points which lie where $P$ is important saturate all the weight in the
averages, and the host of $P$-unimportant points have a negligible relative
weight.  This procedure provides an unbiased estimate but with much noise,
since the actual number of important points employed in the averages is
small. However, if not enough sampling points (sweeps) are used, the Markovian
chain never gets to visit the region where $P$ (but not $P_0$) is
important. In the absence of truly important points, the relative weight of
the $P_0$-distributed points is no longer small, and one obtains a wrong
estimate which is closer to $\esp{A}_{P_0}$ than to $\esp{A}_P$ (assuming the
variation of $w(x)$ can be neglected in the region where $P_0$ is
important). Actually, this expectation is verified by the results displayed in
Table \ref{tab:1} for $\beta=0.5i$ and $8^3$ with $10^5$ sweeps: For the
action in \Eq{1}, $P_0$ corresponds to retaining only $\Re\beta$ in the
action. This is zero in our case and the action becomes ultralocal. An easy
calculation then shows that
\begin{equation}
\esp{O_1}_{P_0} = \esp{I_1}_{P_0} = 0,
\qquad
\esp{O_2}_{P_0} = \esp{I_2}_{P_0} = \beta \esp{\phi^2}_0
\qquad (\Re\beta=0)
.
\end{equation}
The value $\beta \esp{\phi^2}_0= i 117\times 10^{-3}$ is quite consistent with
the RW results quoted in the Table.

This pathology, i.e., the increasing numerical effort required in RW to gain
an overlap with the target distribution, worsens for larger lattices and/or
couplings and prevents us from using RW in those cases. One way to improve the
RW calculation would be to take a better auxiliary distribution $P_0(x)$,
typically, by using effective values of the couplings of quartic and
quadratic operators, and of $\beta$, paralleling the same approach used to
study QCD at finite density \cite{Fodor:2001au}.

Returning to our main focus, the CHB method, we observe that the quality of
the results remains stable for larger lattices ($16^3$ and $8^4$).  Also the
parameter $K$ (we have tried $K=8,10,12$) does not seem to be crucial. Likely,
this is because the error introduced by a finite step $h$ is exponentially
suppressed for boundaryless integrations of smooth functions.

\begin{figure}[t]
\begin{center}
\epsfig{figure=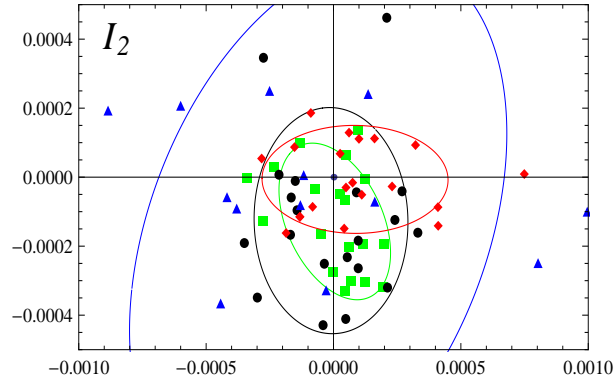,height=50mm,width=80mm}
\end{center}
\caption{Influence of the parameter $Y_s$. The points represent estimates of
$I_2$ from 20 runs with $68\%$ ellipses, for a $8^3$ lattice with $\beta=0.5i$:
  CHB with $Y_s=2$ (green squares), CHB with $Y_s=5$ (black disks), CHB
  with $Y_s=\infty$ (blue triangles), and CL (red rhombuses).}
\label{fig:chb3}
\end{figure}

The CHB results for the observables $O_1$ and $O_2$ are quite stable against
changes in the regulators and their standard deviations are small. On the
other hand the fluctuations are larger in $I_1$ and $I_2$. This follows from
the fact that $O_{1,2}$ are quadratic while $O^\prime_{1,2}$ involve quartic
operators. The worst case displayed in Table \ref{tab:1} corresponds to
removing the regulator $Y_s$, which results in relatively large fluctuations
in $I_{1,2}$ (also, one of the 20 runs aborted). The influence of the value of
$Y_s$ on $\esp{I_2}$ is shown in Fig. \ref{fig:chb3}.

\begin{figure}[t]
\begin{center}
\epsfig{figure=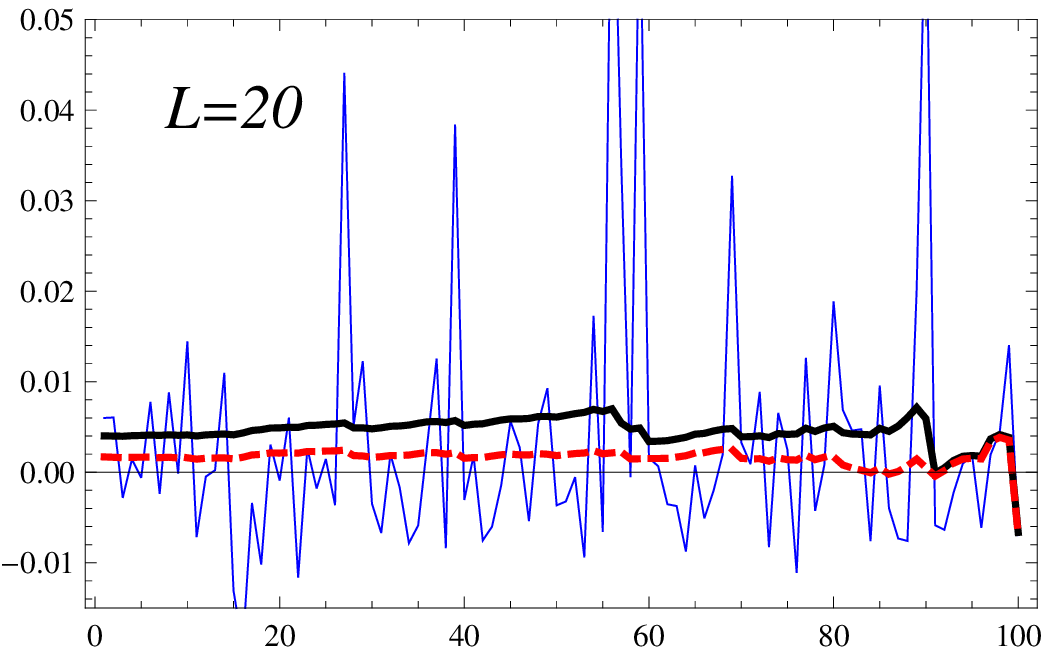,height=50mm,width=80mm}
\epsfig{figure=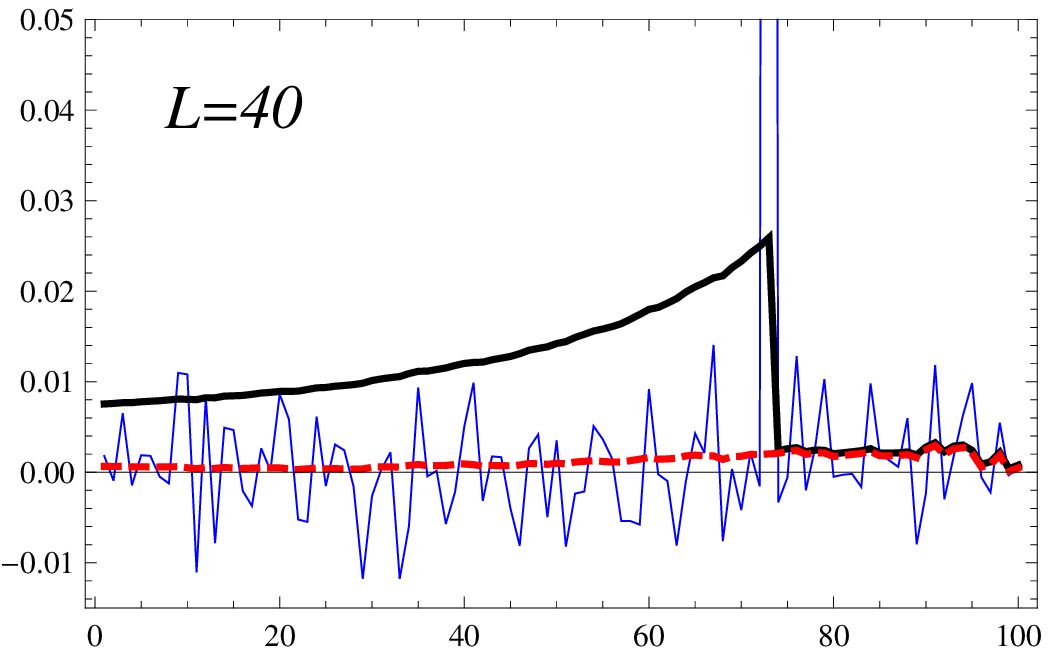,height=50mm,width=80mm}
\end{center}
\caption{Typical runs for $L=20$ and $L=40$ with $\beta=0.5i$ in a $8^3$
  lattice. The $10^5$ sweeps are arranged in $100$ batches of $1000$ sweeps
  each.  The noisy thin solid (blue) line represents $\Re\langle I_1 \rangle$
  for each batch. The thick solid (black) line represents the average over all
  batches to the right (and so computed at later simulation times). The thick
  dashed (red) line shows the effect of removing batches which are off by
  $4\sigma$ ($L=20$) or $8\sigma$ ($L=40$). The high peak in ``$L=40$''
  reaches $0.66$.}
\label{fig:1}
\end{figure}

\begin{figure}[t]
\begin{center}
\epsfig{figure=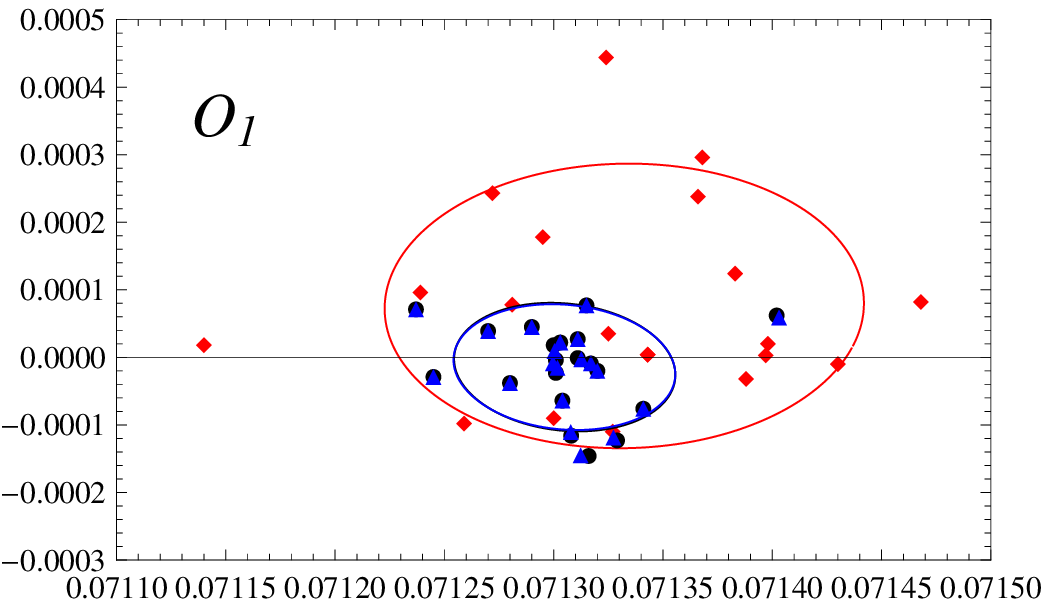,height=50mm,width=80mm}
\epsfig{figure=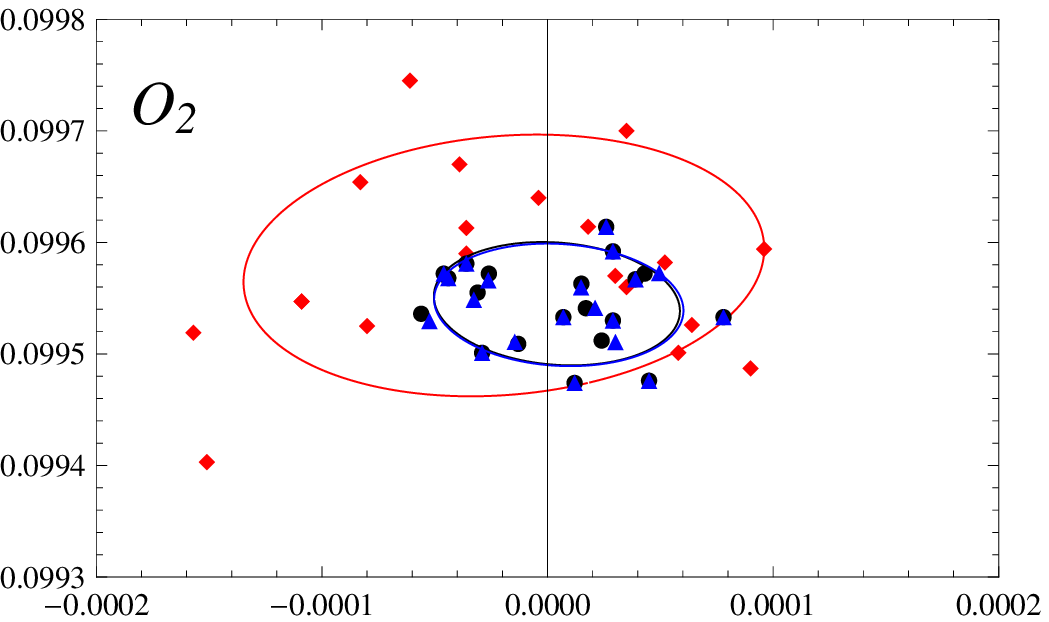,height=50mm,width=80mm}\\
\epsfig{figure=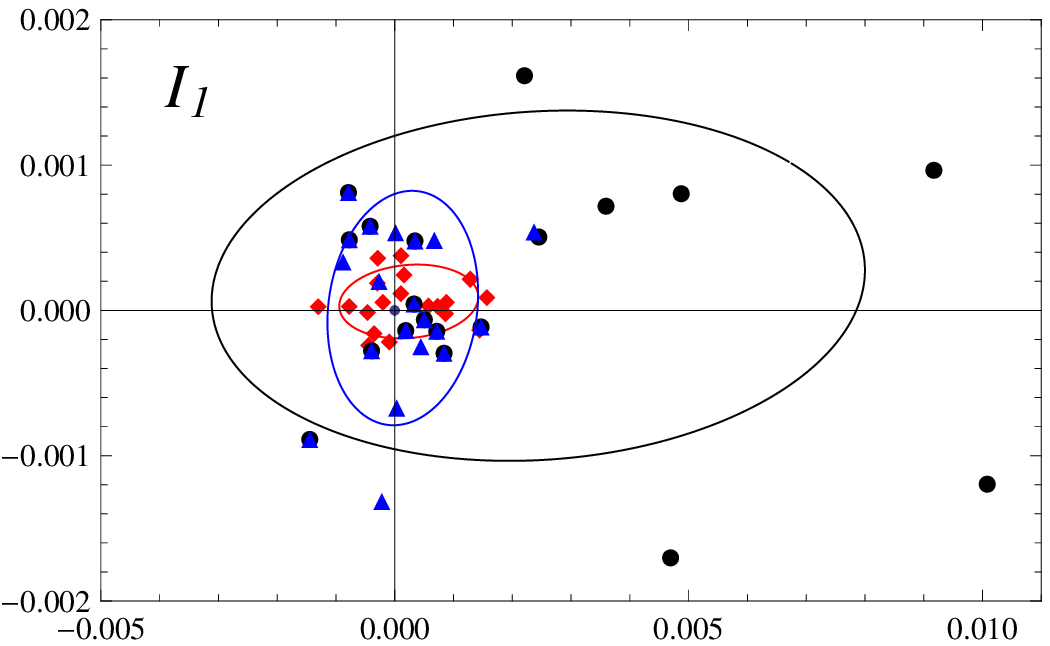,height=50mm,width=80mm}
\epsfig{figure=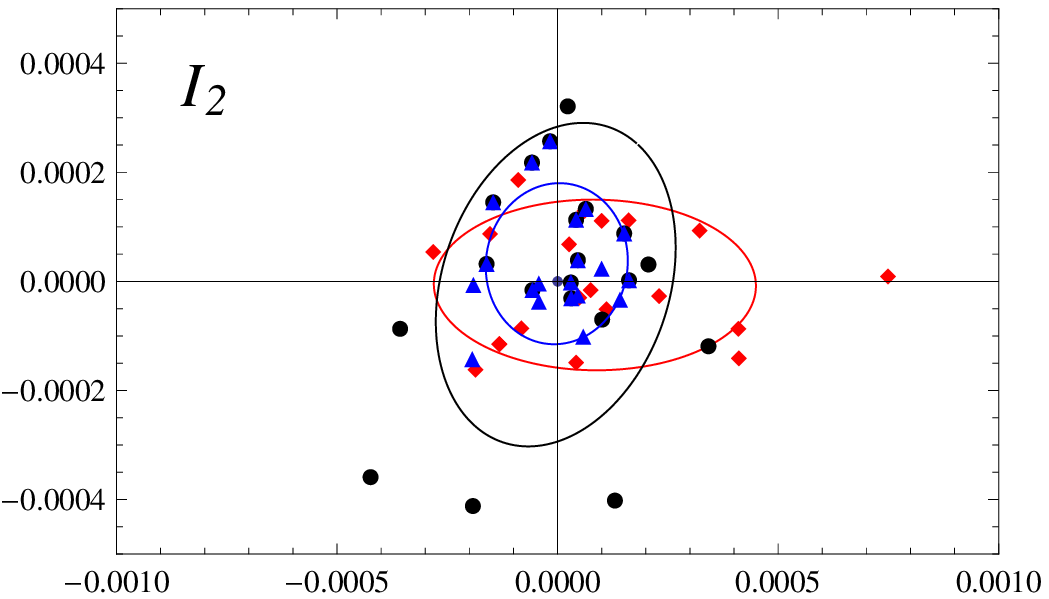,height=50mm,width=80mm}
\end{center}
\caption{Monte Carlo estimates of $O_1$, $O_2$, $I_1$, and $I_2$ for a $8^3$
  lattice with $\beta=0.5i$, using $L=40$ and $K=12$. CL is represented by red
  rhombuses, and CHB by black disks. The blue triangles are from CHB upon
  removal of batches (subsets of 1000 sweeps) lying beyond $8\sigma$ for
  $\Re\esp{I_1}$ (i.e., CHB** in Table \ref{tab:1}). As can be seen, the
  impact of the removal is small on $O_1$ and $O_2$, while $I_1$ and $I_2$ are
  brought closer to zero.  }
\label{fig:chb1}
\end{figure}

A conspicuous feature in CHB for $L=20$ is the small but systematic violation
of $\langle I_1\rangle = 0$. The real part of this observable displays a bias
at the level of $0.005$ that persists for all lattice sizes.\footnote{A
  completely similar bias was observed in CL in preliminary calculations using
  $\Delta t \le 0.01$ and $|\Delta\phi_x|\le 0.01$, together with a noticeably
  shift in $O_1$ and $O_2$. The bias in $I_1$ disappears with the finer steps,
  although one of the $16^3$ runs crashed.} Since the bias is largely reduced
for $L=40$ (to the level of $0.0014$) it would seem that $L=20$ is simply too
small a box for the action of \Eq{1} with $|\beta|=0.5$, however, no bias
exists in similar simulations when $\beta$ is real.\footnote{Also, RW
  calculations putting the system in a similar discretized box show no bias
  (for lattices and couplings for which our RW calculation is reliable).}
Closer inspection reveals that for $L=20$ the various runs have little
dispersion around the biased result. On the contrary, for $L=40$ most runs are
unbiased (at the level of $0.0001$) but for a few of them $ \Re\langle
I_1\rangle $ drifts as far as $0.010$. We show in Fig. \ref{fig:1} a typical
run with $L=20$ and one of the troubled runs with $L=40$. Moderately hard
events are more frequent for the smaller box. For $L=40$, these events become
exceptional and far more violent. Typically, for runs with $Y_s=5$, the ratio
of sites whose update is skipped drops from $3\times 10^{-7}$ for $L=20$ to
less than $10^{-8}$ for $L=40$. As said, in our analysis the runs were
arranged in $100$ batches of $1000$ sweeps each.  For $L=40$, removing batches
in which $\Re\langle I_1\rangle$ is off by $8$ standard deviations introduces
a considerable improvement in the fulfillment of the virial relations without
disturbing $\langle O_{1,2} \rangle$. This can be seen in Table \ref{tab:1},
results marked with CHB**, and also in Fig. \ref{fig:chb1}, where we display
the result of runs with and without removal of eccentric batches. While no
significant difference is appreciated in $O_{1,2}$, the estimates of $I_{1,2}$
are brought closer to zero after the removal.  A quite noteworthy feature,
illustrated in Fig. \ref{fig:1}, is that the Markovian chain quickly recovers
after a troubled batch, leaving no long-lived distortion on the subsequent
batches.

\section{Summary and conclusions}
\label{sec:4}

In this work we have introduced a new approach aimed at the importance
sampling of complex valued distributions through the use of representations.

We start by showing that, quite generally, the more complex (as opposed to
positive definite) a distribution is, the farther into the complex plane must
go any representation of it. Explicit bounds are given on how narrow a strip
parallel to the real axis can be to contain the support of a representation of
a given complex distribution.

Then we develop new techniques to construct representations of complex
probabilities, alternative to the usual complex Langevin. Since the quality of
the representation, as measured by its distance to the real axis, is essential
to have controlled fluctuations in the Monte Carlo estimates, the methods
presented intend to be optimal in this regard. In particular, for the purposes
of illustration, we show representations of complex distributions for which
the complex Langevin approach fails. We also show that the constructions can
be extended for manifolds of dimension higher than one, although with
increasing difficulty.

The specific new proposal for the sampling of complex valued distributions on
manifolds of large dimension, i.e. with many degrees of freedom, is the heat
bath approach. We have not devised a complex version of the Metropolis
algorithm. In its complex version, the Gibbs sampling is implemented by using
a representation of the conditional probability. The formal justification of
the procedure is given in Sec. \ref{sec:3.A}. This is formal since it is not
guaranteed that the chain of updates should converge to a correct
representation of the target complex distribution, it is only shown that such
complex distribution is a fixed point of the algorithm.

To assess the performance of the proposal we consider first a field on a
hypercubic lattice with periodic boundary conditions subjected to a quadratic
action. This makes the update procedure particularly simple. We find that the
method works correctly in this case, provided the imaginary part of the
coupling is not too large.

To do a more thorough analysis we have added a $\lambda\phi^4$ term to the
previous action, thus transforming the problem into a non linear (interacting)
one. The representation of the conditional probability is no longer
straightforward and the new construction methods developed in previous
sections have to be applied. This is needed to achieve representations of
sufficient quality. We show that a Monte Carlo calculation is possible and
obtain various results for different geometries and couplings. Suitable
observables are considered so that generalized virial relations apply. This is
used to check the accuracy of the Monte Carlo results. Also we have compared
with reweighting and with complex Langevin calculations. Due to overlap
problems, our reweighting calculation is only reliable for $3^3$ lattices, and
with more noise for $8^3$ for $\beta=0.25i$.

For $d=1$ lattices, good results are obtained for $\beta$ as large as $i$. For
$d=3,4$, we find that for moderate couplings, $\beta=0.25i$ and $0.5i$, the
observables $O_1$ and $O_2$ and $I_2$ are well reproduced, using reweighting
and/or complex Langevin results and the Schwinger-Dyson relations as
benchmarks. The results tend to deteriorate for larger couplings, as the
random walk makes more frequent visits to regions farther apart from the real
axis.

The quantity $\Re\esp{I_1}$ displays a small systematic bias which depends on
the size of the box used to construct the representations. We have analyzed in
some detail how, for large boxes, this bias is introduced by very specific
contributions which are easy to identify and remove, without altering the
other observables.

Although not free from obstacles, it seems clear that the approach opens
alternative routes for the complex sampling problem, and at least some of the
technical limitations founds could eventually be overcome, in particular,
allowing to treat harder cases (larger $\beta$). In \cite{Salcedo:2007ji} it
is shown that representations exist and can be constructed for complex
distributions defined on compact Lie groups, so there is no impediment of
principle to extend the approach to gauge models.

On a far more hypothetical note, simple variable counting indicates that a
complex distribution can be traded by two positive distributions. We have
illustrated this point with our two-branches representations. In this view,
perhaps it could be possible to replace the standard lattice QCD formulation
with chemical potential, a complex distribution, by a two-branches version,
modeled to be positive, local and having the correct symmetries, as well as a
parameter representing the chemical potential, and letting universality
considerations in the continuum limit to identify it with the standard
formulation. Such approach would avoid altogether the need to deal with
complex distributions and its sampling.

\begin{acknowledgments}
The author thanks C. Garc\'{\i}a-Recio for help with the fast Fourier
transform code, and the Centro de Servicios de Inform\'atica y Redes de
Comunicaciones (CSIRC), Universidad de Granada, for providing computing time.
This work was supported by Spanish Ministerio de Econom\'{\i}a y
Competitividad and European FEDER funds (grant FIS2014-59386-P), and by
the Agencia de Innovaci\'on y Desarrollo de Andaluc\'{\i}a (grant FQM225).
\end{acknowledgments}

\appendix

\end{document}